\documentclass{article}

\usepackage{latexsym}
\usepackage{amsmath}

\makeatletter
\newdimen\normalarrayskip              
\newdimen\minarrayskip                 
\normalarrayskip\baselineskip \minarrayskip\jot
\newif\ifold             \oldtrue            \def\new{\oldfalse}
\def\arraymode{\ifold\relax\else\displaystyle\fi} 
\def\eqnumphantom{\phantom{(\theequation)}}     
\def\@arrayskip{\ifold\baselineskip\z@\lineskip\z@
     \else
     \baselineskip\minarrayskip\lineskip2\minarrayskip\fi}
\def\@arrayclassz{\ifcase \@lastchclass \@acolampacol \or
\@ampacol \or \or \or \@addamp \or
   \@acolampacol \or \@firstampfalse \@acol \fi
\edef\@preamble{\@preamble
  \ifcase \@chnum
     \hfil$\relax\arraymode\@sharp$\hfil
     \or $\relax\arraymode\@sharp$\hfil
     \or \hfil$\relax\arraymode\@sharp$\fi}}
\def\@array[#1]#2{\setbox\@arstrutbox=\hbox{\vrule
     height\arraystretch \ht\strutbox
     depth\arraystretch \dp\strutbox
     width\z@}\@mkpream{#2}\edef\@preamble{\halign
\noexpand\@halignto
\bgroup \tabskip\z@ \@arstrut \@preamble \tabskip\z@ \cr}%
\let\@startpbox\@@startpbox \let\@endpbox\@@endpbox
  \if #1t\vtop \else \if#1b\vbox \else \vcenter \fi\fi
  \bgroup \let\par\relax
  \let\@sharp##\let\protect\relax
  \@arrayskip\@preamble}
\def\eqnarray{\stepcounter{equation}%
              \let\@currentlabel=\theequation
              \global\@eqnswtrue
              \global\@eqcnt\z@
              \tabskip\@centering
              \let\\=\@eqncr

 \halign to \displaywidth\bgroup
    \eqnumphantom\@eqnsel\hskip\@centering
    $\displaystyle \tabskip\z@ {##}$%
    \global\@eqcnt\@ne \hskip 2\arraycolsep
         $\displaystyle\arraymode{##}$\hfil
    \global\@eqcnt\tw@ \hskip 2\arraycolsep
         $\displaystyle\tabskip\z@{##}$\hfil
         \tabskip\@centering
    &{##}\tabskip\z@\cr}
\newfont{\hr}{msbm10}
\newfont{\ams}{msam10}

\hoffset= - 1.0in         

\def\beq{\begin{equation}}
\def\eeq{\end{equation}}
\def\ba{\beq\new\begin{array}{c}}
\def\ea{\end{array}\eeq}
\def\be{\ba}
\def\ee{\ea}

\def\pr{\partial}
\def\bpr{\bar \partial}
\def\lm{\limits}

\title{{\bf Deviation from Alday-Maldacena Duality\\ For Wavy Circle} \vspace{.5cm}}
\author{{\bf Dmitry Galakhov}\footnote{ {\small {\it
MIPT, Moscow, Russia} and {\it ITEP, Moscow, Russia}};
galakhov@itep.ru}, {\bf Hiroshi Itoyama}\footnote{
{\small {\it Osaka City University, Japan, Osaka,
Japan}}; itoyama@sci.osaka-cu.ac.jp}, {\bf Andrei
Mironov}\footnote{ {\small {\it Lebedev Physics Institute} and {\it
ITEP, Moscow, Russia}}; mironov@lpi.ru; mironov@itep.ru} and {\bf
Alexei Morozov}\thanks{{\small {\it ITEP, Moscow, Russia}};
morozov@itep.ru} }

\begin{document}

\maketitle

\vspace{-7.cm}

\begin{center}
\hfill OCU-PHYS-308\\
\hfill FIAN/TD-27/08\\
\hfill ITEP/TH-69/08\\
\end{center}

\vspace{5.5cm}

\begin{abstract}
\noindent Alday-Maldacena conjecture is that the area $A_\Pi$ of
the minimal surface in $AdS_5$ space with a boundary $\Pi$, located in
Euclidean space at infinity of $AdS_5$, coincides with a double
integral $D_\Pi$ along $\Pi$, the Abelian Wilson average in an
auxiliary dual model. The boundary $\Pi$ is a polygon formed by
momenta of $n$ external light-like particles in $N=4$ SYM theory, and in a
certain $n=\infty$ limit it can be substituted by an arbitrary smooth
curve (wavy circle). The Alday-Maldacena conjecture is known to be
violated for $n>5$, when it fails to be supported by the peculiar global
conformal invariance, however, the structure of deviations remains
obscure. The case of wavy lines can appear more convenient for
analysis of these deviations due to the systematic method
developed in
\cite{IMM2} for (perturbative)
evaluation of minimal areas, which is not yet available in the
presence of angles at finite $n$. We correct a mistake in that paper
and explicitly evaluate the $h^2\bar h^2$ terms, where the first
deviation from the Alday-Maldacena duality arises for the wavy circle.
\end{abstract}

\bigskip

\bigskip

\section{Introduction}

Alday-Maldacena hypothesis \cite{AM1}-\cite{AMlast},
an artfully-motivated corollary of the AdS/CFT realization \cite{AdS/CFT}
of string-gauge duality \cite{SGT},
is one of the most spectacular ideas of recent years.
Our understanding of the subject is described in
\cite{MMT1}-\cite{M} and we do not go into any details here.
For us

\bigskip

\centerline{
\framebox{\vbox{\centerline{
the Alday-Maldacena hypothesis is that}\centerline{
the
AdS Plateau problem in its weak form is explicitly resolvable}}}}

\bigskip
\noindent
Moreover, the hypothesis {\it per se} is even more explicit:
the (regularized) area of the minimal surface in AdS space,
\be
A_\Pi = \int \sqrt{\det_{ab} G_{MN}^{AdS}(Y_{min})\partial_a Y_{min}^M
\partial_b Y_{min}^N}
\label{SNG}
\ee
with a given boundary $\Pi$ at the flat infinity (absolute)
of AdS
and
the Abelian Wilson average, the (regularized) double integral along the boundary
\be\label{DLI}
D_\Pi = \frac{1}{4}{\oint\oint}_{\Pi}
\frac{d\vec Y d\vec Y'}{(\vec Y-\vec Y')^2}
\ee
are proportional to each other
\be
\framebox{$
A_\Pi^{reg} \ \stackrel{?}{=}\ \kappa_\Pi D_\Pi^{reg}$}
\label{AMh}
\ee
where $\kappa_\Pi$ depends only on the angles (non-smooth points)
of $\Pi$, not on its smooth deformations.
It is a {\it weak} form of the Plateau problem, because only
the {\it area} of the minimal surface, not its shape
is claimed to be explicitly evaluated.

\bigskip

A physical motivation for this hypothesis comes from string-gauge
duality, supplemented by the BDS conjecture \cite{BDS}:
explicit (hypothetical) formulas for multi-loop diagrams
in the planar limit of $N=4$ Super-Yang-Mills
theory.
The BDS conjecture, adapted in \cite{AM1,dks,BHT} with the help of
the KT-duality \cite{KT}, was that the perturbation
theory at weak coupling provides an answer for the MHV amplitudes
in terms of the Abelian double-loop integral:
\be
{\cal A}_{BDS} \sim \exp (\tilde\kappa D_\Pi)
\label{BDS}
\ee
what allowed to extend it straightforwardly to the strong coupling
regime, by simply changing the coefficient $\tilde \kappa$
to some other $\kappa$.
The $\tilde\kappa$-dependence of the coupling constant is pretty
sophisticated and, as any parameter in the effective action \cite{UFN3},
is controlled by a hidden integrable structure \cite{instr}.
The $\kappa$-$\tilde\kappa$ relation can be also dictated
by this hidden structure, but this still remains to be revealed.

Since at strong coupling the stringy description should be
relevant (with non-critical string described by
the AdS/CFT-correspondence \cite{AdS/CFT}),
the same quantity should be given by $\exp (A_\Pi)$,
and, therefore, one concludes that $A_\Pi = \kappa D_\Pi$,
as conjectured in (\ref{AMh}).

\bigskip

Unfortunately, a closer examination of the BDS conjecture (\ref{BDS})
leads to its serious modification \cite{Vo,Koanti}: as originally
anticipated in \cite{dhks3},
the loop diagrams are actually summed into a {\it non}-Abelian
Wilson average, thus, the {\it answers}  can not be explicitly
exponentiated and continued to the strong coupling
region, which makes the Alday-Maldacena hypothesis (\ref{AMh})
groundless.
Worst of all, the lack of a simple formula like (\ref{BDS})
with momentum-dependence in $\Pi$ and $D_\Pi$
separated from the coupling-constant dependence
in $\tilde \kappa_\Pi$ makes it impossible
to use the ordinary perturbative (diagram) calculations
for analysis of the strong coupling regime:
there is no straightforward way to extrapolate between the
weak and strong coupling regions.
Thus, the relation (\ref{AMh}) and its possible modifications
have to be analyzed directly, without references to diagram
calculations in SYM theory.

This is a well-defined problem, because both sides of
(\ref{AMh}) are pure geometric and contain no reference to
quantum field theory.
However, it is a difficult problem,
because it is difficult to explicitly
construct the minimal AdS surface with a given boundary:
the AdS Plateau problem does not look much simpler than the Euclidean
one, an old "hard problem" in fundamental mathematics.
Currently the number of explicit examples, where the AdS
Plateau problem is {\it fully} resolved
is too small to draw far-going conclusions.
Approximate methods are also difficult to develop, they
are even more involved than in the Euclidean case
because of the need to regularize the area and extract
finite contributions to divergent expressions
(see \cite{IMM1,IM8} for a possible approach, which still
needs to be applied to particular examples).

A way out is provided by consideration of wavy lines \cite{wavy,AM3}:
of small smooth deformations of exactly solvable examples,
i.e. of $\Pi$ deviating slightly and smoothly from a $\Pi_0$,
for which the AdS minimal surface is explicitly known.
This idea is not quite consistent with the original physical
setting, where $\Pi$ were rather polygons with the light-like
sides, and, therefore, are not smooth.
The wavy lines can get relevant in the case of scattering of
$n\rightarrow\infty$ particles with tiny momenta $\sim 1/n$ \cite{AM3}.
Though in this way we loose any direct contact with the finite-$n$
calculations (the only ones that can be made diagrammatically
in the weak coupling regime),
we get instead an infinite-dimensional family of curves $\Pi$
to check if and how eq.(\ref{AMh}) is actually modified.

\section{Alday-Maldacena hypothesis for the wavy circle}

In \cite{IMM2,M} we proposed to do this calculation for
$\Pi_0=$circle,\footnote{
Another possible choice for exactly solvable $\Pi_0$
is a pair of circles \cite{concc}.
This generalization of (\ref{circ}) is very interesting,
because there is a phase transition when the distance
between the two circles large enough as compared with their radii.
}
where the AdS Plateau problem has the explicit solution
\be
r^2 = 1-z\bar z, \ \ \ \ z= y_1+iy_2, \\
y_0=y_3=0
\label{circ}
\ee
Here $y_0,y_1,y_2,y_3,r$ are the standard Poincar\'e coordinates
on $ADS_5$ with the metric
\be
ds^2 = \frac{-dy_0^2 + d\vec y\,^2 + dr^2}{r^2}
\ee
and we refer to \cite{IMM2,M}
for further details about notations and other peculiarities
of our approach.

\bigskip

If the unit circle \ $\Pi_0:\ z=e^{i\phi}$\ is substituted by
any other smooth contour\footnote{
According to Riemann's theorem we can actually parameterize
in this way an {\it arbitrary} smooth planar curve,
not only an {\it infinitesimal} deformation of a circle $\Pi_0$.
}, lying in the plane $y_0=y_3=0$,
\be\label{RiemPar}
\Pi:\ z=H\left(e^{i\phi}\right),\ \ \ \
H(\zeta) = \zeta + \sum_{k=0}^\infty h_k\zeta^k,
\ee
then both sides of eq.(\ref{AMh}) can be represented
as formal series in powers of $h$ and $\bar h$:
\be
\begin{array}{ccccc}
A_\Pi^{reg} &=& A_\Pi - \frac{\pi L_\Pi}{2\mu} + 2\pi &=&
- 3\pi\left(\sum_{m,n} (-)^{m+n}
A^{(m|n)}_{k_1\ldots k_m|l_1\ldots l_n}
h_{k_1+1}\ldots h_{k_m+1}\bar h_{l_1+1}\ldots \bar h_{l_n+1}\right),
\\
D_\Pi^{reg} &=& D_\Pi - \frac{\pi L_\Pi}{4\lambda}
+ \frac{\pi^2}{2} &=&
- \pi^2\left(\sum_{m,n} (-)^{m+n}
D^{(m|n)}_{k_1\ldots k_m| l_1\ldots l_n}
h_{k_1+1}\ldots h_{k_m+1}\bar h_{l_1+1}\ldots h_{l_n+1}\right)
\end{array}
\label{ADwavy}
\ee
In these terms, the Alday-Maldacena hypothesis (\ref{AMh}) is that all
the coefficients $A^{(m|n)}$ and $D^{(m|n)}$ coincide,
$A^{(m|n)} = D^{(m|n)}$, while
\be
\kappa_{smooth}
= \left(\frac{\pi}{3}\right)^{-1}
\ee
Intermediate equations in (\ref{ADwavy}) show
how divergencies are subtracted from the area and the contour integral,
where $\mu$ and $\lambda$ are the corresponding
regularization parameters \cite{IMM2} and
\be
L_\Pi = 2\pi\left(1 + \sum_{m,n} L^{(m|n)}_{k_1\ldots k_m|l_1\ldots l_n}
h_{k_1+1}\ldots h_{k_m+1}\bar h_{l_1+1}\ldots \bar h_{l_n+1}\right)
\label{Lwavy}
\ee
is the length of the contour $\Pi$.
We explicitly write only sums over expansion orders $m$ and $n$,
summations are of course performed also over the indices
$k_i$ and $l_j$.
The unit shift of indices in (\ref{ADwavy}) and (\ref{Lwavy}),
$h_{k+1}$ instead of $h_k$,
is convenient because with such a definition
\be
\sum_{i=1}^m k_i = \sum_{j=1}^n l_j
\label{seru}
\ee
in all the sums (such a shift was not made in \cite{IMM2,M}).

\bigskip

In \cite{IMM2} the first two coefficients were calculated, and they
turned out to be
\be
A^{(1|1)}_{k|k} = \frac{(k+1)k(k-1)}{6} = D^{(1|1)}_{k|k}, \\
A^{(2|1)}_{k_1,k_2|k_1+k_2} = \frac{(k_1+1)(k_2+1)}{12}
\Big(k_1^2+k_2^2+3k_1k_2-k_1-k_2\Big) = D^{(2|1)}_{k_1,k_2|k_1+k_2}
\label{AD1121}
\ee
We explicitly took into account the selection rule (\ref{seru})
and corrected a trivial, but misleading error: in \cite{IMM2,M}
we got a slightly different expression for
$A^{(2|1)}_{k,k|2k}$, what made it different from
$D^{(2|1)}_{k,k|2k}$,
and what we called an {\it anomaly}.
In fact, the entire $A^{(2|1)}$ is what it is in
(\ref{AD1121}),\footnote{
An error in \cite{IMM2} is clear from the line
"$h:=z->h[K]*z^K + h[l]*z^L;$"  in Appendix III
(the very last line at page 21 in the arXiv version of that paper).
It is clear that at $K=L$ there is a double-counting
and the "diagonal" terms with $h_K^2$, $h_K^3$ should be
actually divided by $2^2=4$, $2^3=8$ and so on:
this correction converts $A_\Pi^{reg}$ of \cite{IMM2,M}
into that in (\ref{AD1121}) of the present paper.}
thus $A^{(2|1)} = D^{(2|1)}$
and for the first time {\it the anomaly} shows up in the $(2|2)$ terms,
see below.

Actually (\ref{AD1121}) looks like a serious evidence
{\it in support} of the Alday-Maldacena hypothesis (\ref{AMh}):
it demonstrates the coincidence of {\it infinitely many}
terms at both sides.
As we see below, it looks probable that this
coincidence is not quite accidental: we now believe that it
is extended to all other terms with $m=1$ or $n=1$:
\be
A^{(m|1)}_{k_1,\ldots,k_m|k_1+\ldots+k_m} =
D^{(m|1)}_{k_1,\ldots,k_m|k_1+\ldots+k_m}, \\
A^{(1|n)}_{l_1+\ldots+l_m|l_1,\ldots,l_n} =
D^{(1|n)}_{l_1+\ldots+l_m|l_1,\ldots,l_n}
\label{ADm1}
\ee
(the second relation is of course obtained from the first
one by complex conjugation),
though we check it only up to $m=3$ in this paper.
Moreover,
with additional assumption of polynomial dependence on indices,
(\ref{ADm1}) can be
considered as implication of conformal invariance
\cite{dks,dhks1,dhks3,Komar,IMM2} and already for this reason has good
chances to be true.
At the same time, the conformal invariance does not seem to
control the terms with $min(m,n)\geq 2$ with
indices $k_i+1$ and $l_i+1$ exceeding $2$
(since the invariance algebra contains
only three generators: deformations of $\partial/\partial h_0$,
$\partial/\partial h_1$ and $\partial/\partial h_2$).
Moreover, the index dependence of the area fails to be polynomial in this case
and, therefore, with our present intuition
one can expect deviations from (\ref{AMh}) in the $(2|2)$ terms.
We see below that this expectation turns out to be true:
indeed, already $A^{(2|2)} \neq D^{(2|2)}$ and

\bigskip

\centerline{
\framebox{\vbox{\centerline{
The Alday-Maldacena hypothesis (\ref{AMh}) {\it fails}
for the wavy circle}}}}

\bigskip

\section{The difference between smooth and non-smooth $\Pi$
and non-uni\-ver\-sa\-li\-ty of the $\kappa$-factor}

Of course we already learned from \cite{AM3}
that it fails, but in a different, much weaker sense.
What we know from \cite{AM3} is
that the coefficients $\kappa_\Pi$ can not be fully universal (see the Appendix):
they are different in the $n=4$ case, $\kappa_{\Box} = 1$ \cite{AM1},
and in the $n=\infty$ case with $\Pi =$infinitely long
strip \cite{AM3}, $\kappa_{strip}=4\frac{(2\pi)^2}{\left(\Gamma\left(1/4\right)\right)^4}
\approx 1$.
Moreover, (\ref{ADwavy}) implies \cite{IMM2,M} that
$\kappa_{smooth} = 3/\pi\approx 1$ is different from both
$\kappa_{\Box}$ and $\kappa_{strip}$:
that was another ingredient of {\it the anomaly} of ref.\cite{IMM2}.
However, from (\ref{AD1121}) one can see a possible origin for
this non-universality of the relative coefficient:
the coefficients $A^{(m|n)}$ and $D^{(m|n)}$ are
rational in $k$ and $l$, thus the series
(\ref{ADwavy}) are nicely convergent for smooth
curves $\Pi$ (when $h_k$ falls faster than any negative power
of $k$ at large $k$), but they diverge when $h_k \sim 1/k^2$,
and this is exactly the behavior when angles are present
in $\Pi$.

Indeed, the periodic step function (Fig.\ref{step})
has the Fourier expansion
$$\frac{qh}{\pi^2}\ {\rm Im} \sum_{m=0}^\infty
\frac{e^{iq(2m+1)\phi}}{2m+1},$$
which means that the behavior $h_k \sim \frac{1}{k}$ of
Fourier coefficients can lead to jumps.
The $\phi$-integral of this step function, a saw,
Fig.\ref{pila} has the expansion
$$\frac{h}{\pi ^2}\ {\rm Re} \sum_{m=0}^\infty
\frac{e^{iq(2m+1)\phi}}{(2m+1)^2},$$
thus, $h_k \sim \frac{1}{k^2}$ can provide jumps in the derivative,
i.e. angles in the curve itself.

\begin{figure}
\begin{center}
\begin{picture}(200,40)(50,-20)
\put(10,10){\line(1,0){40}}
\put(50,-10){\line(1,0){40}}
\put(90,10){\line(1,0){40}}
\put(130,-10){\line(1,0){20}}
\put(190,-10){\line(1,0){20}}
\put(210,10){\line(1,0){40}}
\put(250,-10){\line(1,0){40}}
\put(0,0){\line(1,0){150}}
\put(190,0){\vector(1,0){120}}
\put(10,-15){\vector(0,1){40}}
\put(0,-10){\line(1,0){10}}
\put(170,0){\makebox(0,0)[cc]{$\ldots$}}
\put(50,4){\makebox(0,0)[cc]{$\pi/q$}}
\put(90,4){\makebox(0,0)[cc]{$2\pi/q$}}
\put(133,4){\makebox(0,0)[cc]{$3\pi/q$}}
\put(247,4){\makebox(0,0)[cc]{$2\pi-\pi/q$}}
\put(290,4){\makebox(0,0)[cc]{$2\pi$}}
\put(3,11){\makebox(0,0)[cc]{$\frac{qh}{2\pi}$}}
\end{picture}
\caption{{\footnotesize The periodic step function with period $2q$.
The amplitude is adjusted so that the amplitude of saw in
Fig.\ref{pila} is exactly $h$. }}
\label{step}
\end{center}
\end{figure}
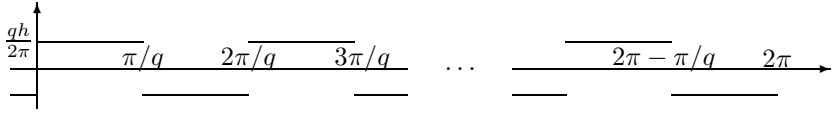
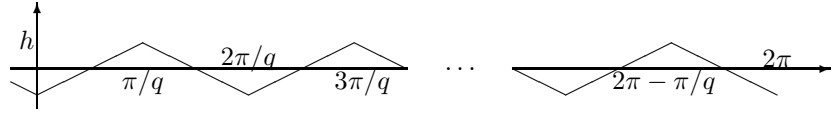
\begin{figure}
\begin{center}
\begin{picture}(200,40)(50,-20)
\put(10,-10){\line(2,1){40}}
\put(50,10){\line(2,-1){40}}
\put(90,-10){\line(2,1){40}}
\put(130,10){\line(2,-1){20}}
\put(190,0){\line(2,-1){20}}
\put(210,-10){\line(2,1){40}}
\put(250,10){\line(2,-1){40}}
\put(0,0){\line(1,0){150}}
\put(190,0){\vector(1,0){120}}
\put(10,-15){\vector(0,1){40}}
\put(0,-5){\line(2,-1){10}}
\put(170,0){\makebox(0,0)[cc]{$\ldots$}}
\put(50,-5){\makebox(0,0)[cc]{$\pi/q$}}
\put(90,4){\makebox(0,0)[cc]{$2\pi/q$}}
\put(133,-5){\makebox(0,0)[cc]{$3\pi/q$}}
\put(247,-5){\makebox(0,0)[cc]{$2\pi-\pi/q$}}
\put(290,4){\makebox(0,0)[cc]{$2\pi$}}
\put(6,11){\makebox(0,0)[cc]{$h$}}
\end{picture}
\caption{{\footnotesize The periodic saw function of period
$2q$ and amplitude $h$.}}
\label{pila}
\end{center}
\end{figure}

Looking already at
the first line in (\ref{AD1121}), one sees that $s=2$
in $h_k \sim \frac{1}{k^s}$ is exactly the convergency boundary
for these series. For the infinite strip of \cite{AM3} with
$$H_{strip}(\zeta) = \log \frac{1-\zeta}{1+\zeta} =
-2\sum_{m=0}^\infty \frac{\zeta^{2m+1}}{2m+1}$$
we have even $h_k \sim \frac{1}{k}$ so that even the length of
$\Pi$,
\be
L_\Pi = 2\pi \left(1 + \frac{1}{4}\sum_{k=0}^\infty k^2|h_k|^2 +
O(h^3)\right)
\ee
diverges, not only $A$ and $D$.

Of course, the regularization of integrals $A_\Pi$ and $D_\Pi$
eliminates these divergencies at the angles and at spatial infinities,
as it does with the divergencies near the boundaries,but the
renormalizations are different for finite terms
(with rapidly decreasing $h_k$), for logarithmic
($h_k\sim 1/k^2$) and for quadratic ($h_k\sim 1/k$) divergencies.
Thus, it is not a big surprise that, if (\ref{AMh}) would even hold
for non-smooth curves, the coefficient $\kappa_\Pi$
can depend on the number of angles.\footnote{
Moreover, one can even imagine that both $A_\Pi^{reg}$ and $D_\Pi^{reg}$
split into two items, one for a smooth, another one for an angle-containing
constituents of the curve $\Pi$, and the two items coincide pairwise but
with different coefficients.}
Evaluation of anomalous ratio $\frac{\kappa_{angles}}{\kappa_{smooth}}
\neq 1$ remains an interesting open problem.

\section{Abelian double integral \label{Dcalc}}

We now proceed to detailed examination of the two sides
of eq.(\ref{AMh}).
The double integral at the r.h.s. in the formula
is actually much simpler than the minimal area in its l.h.s.
In this section, we explicitly represent the {\it generic}
coefficient $D^{(m|n)}$ as a simple multiple sum,
thus, generalizing eq.(1.13) of \cite{IMM2} from $n=1$ or $m=1$
to arbitrary $(m|n)$:
\be
D^{(m;n)}_{k_1,\ldots,k_m| l_1,\ldots,l_n} =
{\rm symmetrized}\left(
\sum_{i_1=0}^{k_1}\ldots \sum_{i_m=0}^{k_m}\
\sum_{j_1=0}^{l_1}\ldots \sum_{j_n=0}^{l_n}\
(k_m-i_m)(l_n-j_n)\right)
= \\ = \frac{1}{p!}\sum_{p=1}^m
\frac{1}{q!}\sum_{q=1}^n \left(
\sum_{i_1=0}^{k_1}\ldots \sum_{i_m=0}^{k_m}\
\sum_{j_1=0}^{l_1}\ldots \sum_{j_n=0}^{l_n}\
(k_p-i_p)(l_q-j_q)\right)
\label{Dgen}
\ee
with the constraint
\be
\sum_{p=1}^m i_p + \sum_{q=1}^n j_q\ =\
\sum_{p=1}^m k_m\ =\ \sum_{q=1}^n l_q
\label{conDgen}
\ee
We derive this formula in subsection
\ref{derivD}, but consider a number
of examples and applications before.

\subsection{Examples}

In particular, (\ref{AD1121}) is an immediate
corollary of (\ref{Dgen}):
\be
D^{(2)} \equiv \sum_{\stackrel{k,l=0}{k=l}}^\infty
D^{(1|1)}_{k|l} h_{k+1}\bar h_{l+1}
= \sum_{k=0}^\infty |h_{k+1}|^2
\left.\sum_{i=0}^k (k-i)(k-j)\right|_{i+j=k}
= \sum_{k=0}^\infty \frac{(k-1)k(k+1)}{6}|h_{k+1}|^2
\ee
\be
D^{(3)} \equiv \sum_{\stackrel{k_1,k_2,l=0}{k_1+k_2=l}}^\infty
D^{(2|1)}_{k_1,k_2|l} h_{k_1+1}h_{k_2+1}\bar h_{l+1}
+ \sum_{\stackrel{k,l_1,l_2=0}{k=l_1+l_2}}^\infty
D^{(1|2)}_{k|l_1,l_2} h_{k+1}\bar h_{l_1+1}\bar h_{l_2+1}
\ee
with
\be
D^{(2|1)}_{k_1,k_2|k_1+k_2}  = {\rm symm}_{k_1\leftrightarrow k_2}
\left(
\left.\sum_{i_1=0}^{k_1}\sum_{i_2=0}^{k_2}
\sum_{j=0}^{k_1+k_2} (k_2-i_2)(k_1+k_2-j)\right|_{i_1+i_2+j=k_1+k_2}
\right) = \\ = \frac{1}{12}(k_1+1)(k_2+1)
(k_1^2+3k_1k_2+k_2^2-k_1-k_2)
= D^{(1;2)}_{k_1+k_2;k_1,k_2}
\ee
Further, one easily reproduces from (\ref{Dgen}) the result of \cite{IMM2} for
more terms
\be
D^{(4)} \equiv
\sum_{\stackrel{k_1,k_2,k_3,l=0}{k_1+k_2+k_3=l}}^\infty
D^{(3|1)}_{k_1,k_2,k_3|l} h_{k_1+1}h_{k_2+1}h_{k_3+1}\bar h_{l+1}
+ \\
+ \sum_{\stackrel{k,l_1,l_2=0}{k_1+k_2=l_1+l_2}}^\infty
D^{(2|2)}_{k_1,k_2|l_1,l_2} h_{k_1+1}h_{k_2+1}
\bar h_{l_1+1}\bar h_{l_2+1}
+ \sum_{\stackrel{k,l_1,l_2,l_3=0}{k=l_1+l_2+l_3}}^\infty
D^{(1|3)}_{k|l_1,l_2,l_3} h_{k+1}\bar h_{l_1+1}\bar h_{l_2+1}
\bar h_{l_3+1}
\ee
with
\be
D^{(3|1)}_{k_1,k_2,k_3|k_1+k_2+k_3}  =
{\rm symm}_{k_1,k_2,k_3}
\left(
\left.\sum_{i_1=0}^{k_1}\sum_{i_2=0}^{k_2}\sum_{i_3=0}^{k_3}
\sum_{j=0}^{k_1+k_2+k_3} (k_3-i_3)(k_1+k_2+k_3-j)
\right|_{i_1+i_2+i_3+j=k_1+k_2+k_3}
\right) = \\ = \frac{1}{18}(k_1+1)(k_2+1)(k_3+1)
\Big(k_1^2+k_2^2+k_3^2 + 3(k_1k_2 + k_2k_3 + k_3k_1)-k_1-k_2-k_3\Big)
= D^{(1|3)}_{k_1+k_2+k_3|k_1,k_2,k_3}
\ee
and
\be
D^{(2;2)}_{k_1,k_2|l_1,l_2} =
{\rm symm}_{k_1\leftrightarrow k_2}
\ {\rm symm}_{l_1\leftrightarrow l_2}
\left(
\left.\sum_{i_1=0}^{k_1}\sum_{i_2=0}^{k_2}\sum_{j_1=0}^{l_1}
\sum_{j_2=0}^{l_2} (k_2-i_2)(l_2-j_2)
\right|_{i_1+i_2+j_1 + j_2=k_1+k_2=l_1+l_2}
\right)
\ee

\subsection{Generating functions}

To more effectively deal with various double integral formulas, one can deal, instead of the
coefficients $D$'s, with the generating functions. To this end, one introduces the
following generating functions:
\be
D(x_1...x_m;y_1...y_n)\equiv \sum D^{(m|n)}_{k_1...k_m|l_1...l_n}x_1^{k_1}...x_m^{k_m}
y_1^{l_1}...y_m^{l_m}\\
h(x)\equiv \sum h_nx^n\\
\bar h(x)\equiv\sum\bar h_nx^n
\ee
Then, for instance\footnote{Note that with another natural definition of generating functions
$D$'s,
\be
D(x_1...x_m;y_1...y_n)\equiv \sum {D^{(m|n)}_{k_1...k_m|l_1...l_n}\over x_1^{k_1+2}...x_m^{k_m+2}
y_1^{l_1+2}...y_m^{l_m+2}}
\ee
formula (\ref{gfDh}) looks even simpler:
\be
D^{(m,1)}=\sum D^{(m,1)}_{i_1\ldots i_m}h_{i_1+1}\ldots h_{i_m+1} \bar h_{i_1+\ldots+i_m+1}=
\oint\ldots\oint D^{(m,1)}(x_1,\ldots,x_m;y)h(x_1)\ldots h(x_m)
\bar h(y)\frac{dx_1}{2\pi i}\ldots \frac{dx_m}{2\pi i}\frac{dy}{2\pi i}
\ee
with $h(x)$ instead of $h(\bar x)$ and the integration contours encircling zero, but not
obligatory unit circles.},
\be\label{gfDh}
D^{(m,1)}=\sum D^{(m,1)}_{i_1\ldots i_m}h_{i_1+1}\ldots h_{i_m+1} \bar h_{i_1+\ldots+i_m+1}=\\
=\oint\ldots\oint D^{(m,1)}(x_1,\ldots,x_m;y)h(\bar x_1)\ldots h(\bar x_m)
\bar h(\bar y)\frac{dx_1}{2\pi i}\ldots \frac{dx_m}{2\pi i}\frac{dy}{2\pi i}
\ee
where integration goes over unit circles, and similarly for other $D$'s.

\subsubsection{The simplest generating functions}

The simplest generating functions can be immediately obtained from explicit formulas in indices:
\be
D^{(1,1)}_{i} = \frac{i+1}{6}(i^2-i), \\
D^{(2,1)}_{ij} = \frac{(i+1)(j+1)}{12}(i^2+j^2+3ij - i - j),\\
D^{(3,1)}_{ijk} = \frac{(i+1)(j+1)(k+1)}{18}
\Big(i^2+j^2+k^2 + 3(ij+jk+ki) - i - j - k\Big), \\
\ldots  \\
D^{(m,1)}_{i_1\ldots i_m} =
\frac{(i_1+1)\ldots(i_m+1)}{6m}\left(
\sum_{p=1}^m i_p^2 + 3\sum_{p<q} i_pi_q - \sum_{p=1} i_p\right)
\ee
These series can be summed up to give the following expressions
\be
D^{(1;1)}(x,y) = \frac{(xy)^2}{(1-xy)^4}
= \Big(\xi(1 -\xi)\Big)^2,  \\
D^{(2,1)}(x_1,x_2;y) = \frac{(x_1y+x_2y - 2x_1x_2y^2)^2}
{2(1-x_1y)^4(1-x_2y)^4} =
\frac{1}{2}\Big(\xi_1\xi_2(2 - \xi_1-\xi_2)\Big)^2,  \\
\ldots \\
D^{(m,1)}(x_1,\ldots,x_m;y) =
\frac{1}{m}\Big(\xi_1\ldots \xi_m(m - \xi_1 - \ldots - \xi_m)\Big)^2
\ee
where $\xi_i = \frac{1}{1-x_iy}$.

\subsubsection{Generating functions and Miwa variables}

Another possibility, which allows one to look at the generating functions differently, is
to introduce instead of the variables $h_k$ and $\bar h_k$ (we omit from these sets
$h_0$ and $\bar h_0$ since they anyway do not enter the results due to the conformal
invariance) the two infinite sets of new
Miwa-like variables $\{x_i\}$ and $\{y_i\}$ \cite{Miwa,UFN3}
\be
h_n\equiv\sum_i \alpha_i x_i^{n-1}\ ,\ \ \ \ \ \ \ \ \ \ \bar h_n\equiv\sum_i \bar\alpha_i y_i^n
\ee
where $\alpha_i$ and $\bar\alpha_i$ are multiplicities, i.e. the number of coinciding
$x_i$ and $y_i$ correspondingly. Then,
\be
z(\zeta)=\zeta+\sum_{k=2} h_k\zeta^k=\zeta+\sum\lm_{i,k}\alpha_ix_i^k \zeta^{k+1}=
\zeta+\sum_i\frac{\alpha_i\zeta}{1-x_i\zeta}
\ee
and similarly
\be
\bar z(\zeta)=\bar\zeta+\sum\lm_j\frac{\bar\alpha_j\bar\zeta}{1-y_i\bar\zeta}\\
\ee
In fact, one can generically put all the multiplicities equal to
unity. However, one can instead preserve non-trivial multiplicities in order to immediately obtain
the following formula that allows one to produce the generating functions from the double loop
integral $D[z,\bar z]$
\be
D^{(n|m)}(x|y)=\left.\left[\left(\prod\lm_{i=1}^n\frac{\pr}{\pr \alpha_i}
\prod\lm_{j=1}^m\frac{\pr}{\pr \bar\alpha_j}\right)D[z,\bar z]\right]\right|_{\alpha_i,\beta_j=0}
\ee
The derivatives here should be taken w.r.t. different $\alpha_i$'s and, similarly, w.r.t.
different $\bar\alpha_i$'s.

\subsubsection{Generating functions and Schwarzian}

It turns out that all the terms $D^{(m|1)}$ can be explicitly summed up
\cite{IMM2} or derived by the above described technique, and are expressed through the
Schwarzian derivative:
\be
D^{(\cdot|1)} \equiv
\sum_{m=1}^\infty D^{(m|1)}_{k_1\ldots k_m|k_1+\ldots+k_m}
h_{k_1+1}\ldots h_{k_m+1}\bar h_{k_1+\ldots+k_m+1} =-\frac{1}{6}
\oint \bar h(\bar \zeta)S_\zeta(z)\zeta^2 d\zeta
\ee
where the contour integral is taken along the unit circle
$\zeta = e^{i\phi}$, $z = \zeta + \sum_k h_k\zeta^k$ and the Schwarzian derivative is
\be
S_\zeta(z) = \frac{z'''}{z'} - \frac{3}{2}\left(\frac{z''}{z'}\right)^2
\ee
Similarly one can simply derive that
\be
D(h,\bar h)=\pi^2\sum\lm_{n=1}^{\infty}(-1)^{n}\left\{\frac{1}{2\pi i}
\oint d\zeta_k \bar h(\bar \zeta_k)\right\}_{k=1}^n
\left(\frac{1}{6}\sum_{i}\frac{\zeta_i^{2n}}{\prod\lm_{j\neq i}(\zeta_i-\zeta_j)^2}S(z_i)-\sum\lm_{i\neq j}\frac{2\zeta_i^n \zeta_j^n \sigma(z_1,z_2)}{(\zeta_i-\zeta_j)^2\prod\lm_{p\neq i,j}(\zeta_i-\zeta_p)(\zeta_j-\zeta_p)}\right)
\label{hexp}
\ee
where
\be
\sigma(z_1,z_2)=\frac{z_1' z_2'}{(z_1-z_2)^2}
\ee
Advantage of the representation is that $s$ and $\sigma$ are explicit invariants of the conformal
symmetry, see s.6.3 below.

\subsection{Derivation of (\ref{Dgen}) \label{derivD}}

To complete this section, we now return to the formal derivation of (\ref{Dgen}):

$$
D_\Pi = \oint_\Pi\oint_\Pi \frac{\frac{1}{2}(dzd\bar z' + d\bar zdz')}
{|z-z'|^2}
= \frac{1}{2} \oint_{|\zeta|^2=1} \oint_{|\zeta'|^2=1}
PQ\bar Q\ \frac{d\zeta d\bar\zeta'}
{|\zeta-\zeta'|^2} + (\zeta\leftrightarrow \zeta',
\bar\zeta\leftrightarrow \bar\zeta'),
$$
where
$$
P = 1 + \frac{\partial h(\zeta)}{\partial\zeta} +
\frac{\overline{\partial h(\zeta')}}{\partial\bar\zeta'} +
\frac{\partial h(\zeta)}{\partial\zeta}
\frac{\overline{\partial h(\zeta')}}{\partial\bar\zeta'} =(1 + p_{1|0})(1 + p_{0|1})=
1 + p_{1|0} + p_{0|1} + p_{1|1}
$$
with $p_{1|1} = p_{1|0}p_{0|1}$, while
$$
Q = \left(1+\frac{h(\zeta)-h(\zeta')}{\zeta-\zeta'}\right)^{-1}
= (1+q_{1|0})^{-1}
$$
and $\bar Q = (1+ q_{0|1})^{-1}$ with $q_{0|1}=\overline{q_{1|0}}$.
The subscripts here count the numbers of $h$ and $\bar h$
in the corresponding expressions.

It is easy to check\footnote{
Indeed, the direct computation gives:
$$
(PQ\bar Q)_{0|0} = 1;
$$
$$
(PQ\bar Q)_{1|0} = p_{1|0} - q_{1|0},\ \ \ \
(PQ\bar Q)_{0|1} = p_{0|1} - q_{0|1};
$$
$$
(PQ\bar Q)_{2|0} = p_{1|0}(-q_{1|0}) + (-q_{1|0})^2 =
(-q_{1|0})(p_{1|0} - q_{1|0}), \ \ \ \
(PQ\bar Q)_{0|2} = p_{0|1}(-q_{0|1}) + (-q_{0|1})^2 =
(-q_{0|1})(p_{0|1} - q_{0|1}),
$$
$$
(PQ\bar Q)_{1|1} = p_{1|1} - p_{1|0}q_{1|0} - p_{0|1}q_{0|1} +
q_{1|0}q_{0|1} = (p_{1|0}-q_{1|0})(p_{0|1}-q_{0|1});
$$
and so on. Eq.(\ref{PQQ}) is an obvious generalization of
elementary formulas, obtained in this straightforward way.
}
that for all $m,n\geq 1$
\be
(PQ\bar Q)_{m|n} = (-q_{1|0})^{m-1}(-q_{0|1})^{n-1}
(p_{1|0}-q_{1|0})(p_{0|1}-q_{0|1})
\label{PQQ}
\ee
Now,
$$
p_{1|0}-q_{1|0} = \frac{\partial h(\zeta)}{\partial\zeta} -
\frac{h(\zeta)-h(\zeta')}{\zeta-\zeta'} =
\sum_{k=1}^\infty h_k\zeta^{k-1}
\left(k - \frac{1-(\zeta'/\zeta)^k}{1-\zeta'/\zeta}\right)
$$
If it is further divided by $(1-\zeta'/\zeta)$, then
\be
\frac{p_{1|0}-q_{1|0}}{1-\zeta/\zeta'}\ =\
\sum_{k=1}^\infty h_k\zeta^{k-1}\,
\frac{k-1-x-\ldots - x^{k-1}}{1-x}\ =  \\ =
\sum_{k=1}^\infty h_k\zeta^{k-1}\Big(
1 + (1+x) + (1+x+x^2) + \ldots + (1+x+\ldots +x^{k-2})\Big) = \\
= \sum_{k =1}^\infty\sum_{i=0}^{k-1} (k-i-1) h_k\zeta^{k-1}
(\zeta'/\zeta)^i
\ee
where $x=\zeta'/\zeta$.
Similarly,
\be
\frac{p_{0|1}-q_{0|1}}{1-\bar\zeta/\bar\zeta'} =
\sum_{l=1}^\infty \sum_{j=0}^{l-1} (l-j-1)\bar h_l(\bar\zeta')^{l-1}
(\bar\zeta/\bar\zeta')^j
\ee
At the same time
\be
q_{1|0} = \frac{h(\zeta)-h(\zeta')}{\zeta-\zeta'} =
\sum_{k=1}^\infty h_k\zeta^{k-1} \sum_{i=0}^{k-1} (\zeta'/\zeta)^i
\ee
and
\be
q_{0|1} = \frac{h(\bar\zeta)-h(\bar\zeta')}{\bar\zeta-\bar\zeta'} =
\sum_{l=1}^\infty \bar h_l(\bar\zeta')^{l-1} \sum_{j=0}^{l-1}
(\bar\zeta/\bar\zeta')^j
\ee
It now follows from (\ref{PQQ}) that
$$
\frac{(PQ\bar Q)_{m|n}}{(1-\zeta'/\zeta)(1-\bar\zeta/\bar\zeta')}
= (-)^{m+n}\left(\sum_{k=1}^\infty h_k\zeta^{k-1}
\sum_{i=0}^{k-1} (\zeta'/\zeta)^i\right)^{m-1}\!\!
\left(\sum_{k =1}^\infty\sum_{i=0}^k (k-i) h_k\zeta^{k-1}
(\zeta'/\zeta)^i\right)
\cdot $$ $$ \ \ \ \ \ \ \ \ \ \ \ \ \ \ \ \ \ \ \ \ \ \ \cdot
\left(\sum_{l=1}^\infty \bar h_l(\bar\zeta')^{l-1} \sum_{j=0}^{l-1}
(\bar\zeta/\bar\zeta')^j\right)^{n-1}\!\!
\left(\sum_{l=1}^\infty \sum_{j=0}^l (l-j)\bar h_l(\bar\zeta')^{l-1}
(\bar\zeta/\bar\zeta')^j\right) =
$$ $$
= (-)^{m+n}\sum_{k_1,\ldots,k_m=0}^\infty\
\sum_{l_1,\ldots,l_n=0}^\infty
h_{k_1+1}\ldots h_{k_m+1}\bar h_{l_1+1}\ldots \bar h_{l_n+1}
\cdot $$ $$ \cdot
\zeta^{\sum_{p=1}^m k_p}(\bar\zeta')^{\sum_{q=1}^n l_q}
\prod_{p=1}^{m-1} \left(\sum_{i=0}^{k_p} (\zeta'/\zeta)^i\right)
\left(\sum_{i=0}^{k_m} (k_m\!-i) (\zeta'/\zeta)^i\right)
\prod_{q=1}^{n-1} \left(\sum_{j=0}^{l_q} (\bar\zeta/\bar\zeta')^i\right)
\left(\sum_{j=0}^{l_n} (l_n\!-j) (\bar\zeta/\bar\zeta')^j\right)
= $$ $$ = (-)^{m+n}\sum_{k_1,\ldots,k_m=0}^\infty\
\sum_{l_1,\ldots,l_n=0}^\infty
h_{k_1+1}\ldots h_{k_m+1}\bar h_{l_1+1}\ldots \bar h_{l_n+1}
\cdot $$ $$ \cdot
\zeta^{\sum_{p=1}^m k_p}(\bar\zeta')^{\sum_{q=1}^n l_q}
\sum_{i_1=0}^{k_1}\ldots \sum_{i_m=0}^{k_m}
\sum_{j_1=0}^{l_1}\ldots \sum_{j_n=0}^{l_n}
(\zeta'/\zeta)^{\sum_{p=1}^m i_p}
(\bar\zeta/\bar\zeta')^{\sum_{q=1}^n j_p}
(k_m\!-i_m)(l_n\!-j_n)
$$
Once again one can see convenience of the shift of indices $k\rightarrow k+1$
and $l\rightarrow l+1$ in $h_k$ and $\bar h_l$.

Integration over $d\zeta/\zeta$ and $d\bar\zeta'/\bar\zeta'$
along the unit circle now imposes the constraints (\ref{conDgen})
\be
\sum_{p=1}^m i_p + \sum_{q=1}^n j_q = \sum_{p=1}^m k_p +
\sum_{q=1}^n l_q
\ee
and provides an expression (\ref{Dgen}) for generic
\be
D^{(m|n)}_{k_1\ldots k_m|l_1\ldots l_n} =
{\rm symmetrized}\left( \sum_{i_1=0}^{k_1}\ldots \sum_{i_m=0}^{k_m}
\sum_{j_1=0}^{l_1}\ldots \sum_{j_n=0}^{l_n}(k_m\!-i_m)(l_n\!-j_n)\right)
\ee

\section{Minimal area \label{Acalc}}

Turning to the l.h.s. of eq.(\ref{AMh}),
we begin with briefly reminding the perturbation theory
for AdS minimal surfaces, suggested in
\cite{IMM1,IM8} and adjusted to the wavy lines problem
in \cite{IMM1}.

\subsection{Perturbative evaluation of the minimal surface}

After substitution of
\be
y_0 = 0, \\
z= y_1+iy_2 = \zeta + H(\zeta) = \zeta + \sum_{k\geq 0}^\infty
h_k\zeta^k, \\
r(\zeta,\bar\zeta) = \sqrt{1-\zeta\bar\zeta + a(\zeta,\bar\zeta)}
\ee
the AdS Nambu-Goto action (\ref{SNG})
becomes
\be\label{plan-NGA}
A_\Pi = \int\frac{\sqrt{|\partial H|^2
\Big(|\partial H|^2r^2 + |\partial r^2|^2\Big)}}
{r(r^2+\mu^2)}\ d^2\zeta \ \stackrel{\mu = 0}{\longrightarrow} \
\int \frac{|1+\partial h|^2 \sqrt{1-\zeta\bar\zeta + a +
\frac{(\partial a - \bar\zeta)(\bar\partial a - \zeta)}{|1+\partial h|^2}}}
{(1-\zeta\bar\zeta +a)^{3/2}}\ d^2\zeta
\equiv \int {\cal L}d^2\zeta
\ee
where solution $a(\zeta,\bar\zeta)$ for the equations of motion,
i.e. the shape of the minimal surface should be substituted,
subjected to the boundary $a(e^{i\phi},e^{-i\phi}) = 0$
at $\Pi$, and $\mu$ is a small IR-regularization parameter
(one can ignore it in solving the equations of motion, but
it is important for evaluating the minimal area, i.e. the action
itself).

This action implies the equation of motion in the form
\be
0 =
\partial\left(\frac{\partial {\cal L}}{\partial (\partial a)}\right)
+ \bar\partial\left(\frac{\partial {\cal L}}
{\partial (\bar\partial a)}\right) -
\frac{\partial{\cal L}}{\partial a} =
\frac{1}{4(1-\zeta\bar\zeta)^{3/2}}\left\{
\Delta_{NG}\Big(a(\zeta,\bar\zeta)\Big)
+ R(a;h,\bar h)\right\}
\ee
with the Nambu-Goto operator
\be
\Delta_{NG} = 4\partial\bar\partial
- \zeta^2\partial^2-2\zeta\bar\zeta\partial\bar\partial -
\bar\zeta^2\bar\partial^2
\label{NGO}
\ee
(it depends on the AdS metric and on the choice of the
circle solution $r_0^2 = 1-\zeta\bar\zeta$)
and with a somewhat sophisticated,
but straightforwardly derivable
expression $R(a;h,\bar h)$.

The equation of motion can be now iteratively expanded in powers
of $h$ and $\bar h$, as we did before in study of the double
contour integral:
\be\label{NGE}
\Delta_{NG}\Big(a^{(1)}(\zeta,\bar\zeta)
+ \bar\zeta h(\zeta) + \zeta\bar h(\bar\zeta)\Big) = 0, \\
\Delta_{NG}\Big(a^{(k)}(\zeta,\bar\zeta)
\Big) =
- R^{(k)}(a;h,\bar h) \ \ \ \ {\rm for} \ k\geq 2.
\ee
The superscripts here refer to the order in $h$, and
we used here the fact \cite{IMM2} that $R^{(0)} = R^{(1)} = 0$.
Since $R(a;h,\bar h)$ is a non-linear function
of all its arguments, all the components
$a^{(j)}$ with $1\leq j \leq k-1$ contribute to $R^{(k)}$.

The $h^k$-component $a^{(k)}$  contributes only to the
$h^{2k}$-component of the action, $A^{(2k)}$, since
$a^{(1)}+\ldots+a^{(k-1)}$ satisfies the equation of motion:
because of this, the $a^{(k)}$-linear term in
$$A\Big(a^{(1)}+\ldots+a^{(k-1)}+a^{(k)}\Big) =
A\left(a^{(1)}+\ldots+a^{(k-1)}\right)  +
a^{(k)}\frac{\delta A}{\delta a}\left(a^{(1)}+\ldots+a^{(k-1)}\right)
+ \ldots$$
is vanishing, and $a^{(k)}$ can contribute only
quadratically, i.e. only to $A^{(2k)}$.
In particular, $A^{(2)}$ and $A^{(3)}$ depend only on $a^{(1)}$
\cite{IMM2}, into $A^{(4)}$ and $A^{(5)}$ only $a^{(1)}$ and $a^{(2)}$
are contributing, $a^{(3)}$ first contributes to $A^{(6)}$ and so on.

It can be instructive to look at the polynomial example here.
Let us take for the action
$$
S(a,h) = -ah + \frac{1}{2}a^2\Big(1-\alpha h - \beta h^2 - \gamma h^3
+ \ldots\Big)
$$
Then, if one substitutes a solution of the equations of motion
expanded in powers of $h$,
$$
a = a_1 + a_2 + a_3 + a_4 + \ldots,
$$
one obtains
$$
S_{min} = \left(\frac{1}{2}a_1^2-a_1h\right)
+ \left(a_1a_2-\alpha a_1^2 h - a_2h\right)
+ \left(\frac{1}{2}a_2^2 + a_1a_3 - \alpha a_1a_2h
-\frac{1}{2}\beta a_1^2 h^2 - a_3h\right) + $$ $$
+ \left(a_2a_3 + a_1a_4 - \frac{1}{2}a_2^2 h - \alpha a_1a_2h
- \beta a_1a_2 h^2 - \frac{1}{2}\gamma a_1^2h^3 - a_4h\right)
+ O(h^6)
$$
The coefficients in front of $a_3$ and $a_4$ in this expressions
are respectively $(a_1-h) + (a_2-\alpha a_1h)$ and $(a_1-h)$.
However, these brackets are vanishing by definition of $a_1$ and $a_2$
which are the iterative solutions to the equations of motion
$$
a = \frac{h}{1-\alpha h - \beta h^2 - \gamma h^3 + \ldots}
= h + \alpha h^2 + (\beta+\alpha^2)h^3
+ (\gamma+2\alpha\beta+\alpha^3)h^4 + \ldots
$$

\subsection{Inverting Nambu-Goto operator}

Now we need a systematic way to solve the Nambu-Goto equations
with the non-vanishing right hand sides and with the
vanishing (Dirichlet) boundary conditions.

The zero modes of Nambu-Goto operator $\Delta_{NG}$, eq.(\ref{NGO})
are defined through
\be
g_k(\eta) = \frac{1+k\sqrt{1-\eta}}{(1+\sqrt{1-\eta})^k},
\ \ \ \ \ \
\tilde g_k(\eta) = \frac{1-k\sqrt{1-\eta}}{(1-\sqrt{1-\eta})^k}
= \frac{(1-k\sqrt{1-\eta})(1+\sqrt{1-\eta})^k}{\eta^k},\\
G_k(u) = g_k(1-u^2) = \frac{1+ku}{(1+u)^k}, \ \ \ \
\tilde G_k(u) = \tilde g_k(1-u^2) = \frac{1-ku}{(1-u)^k},\ \ \ \
k\geq 2
\ee
with $\eta=\zeta\bar\zeta$ and $u=\sqrt{1-\eta} = \sqrt{1-\zeta\bar\zeta}$
as follows:
\be
\Delta_{NG} \Big(\zeta^k g_k(\eta)\Big) = 0, \ \ \ \
\Delta_{NG} \Big(\zeta^k \tilde g_k(\eta)\Big) = 0
\ee
The solution to
\be
\Delta_{NG}\Big(\zeta^k F_k(\eta)\Big) = \zeta^k m_k(\eta),
\ee
which is finite at $\eta=0$ and satisfies $F_k(\eta=1)=0$,
is provided by the usual "variation-of-constants" method
\be
F_k(\eta) = -\tilde g_k(\eta) \int^{1}_{\sqrt{1-\eta}}
\frac{M_k(u)d\mu_k}{\tilde G_k(u)} +
g_k(\eta) \left\{\int^{1}_{\sqrt{1-\eta}}
\frac{M_k(u)d\mu_k}{G_k(u)} +
\int_0^1 \left(\frac{1}{\tilde G_k(u)}-\frac{1}{G_k(u)}\right)
M_k(u)d\mu_k\right\},
\ee
where $M_k(u) = m_k(1-u^2)$,
\be
d\mu_k=\frac{(1-k^2 u^2)du}{2k(k^2-1)u^2}
\ee
and each of the three integrals is convergent for
non-singular $M_k(\eta)$.
In the particular cases of $k=0$ and $k=1$,
these formulas are not directly applicable,
we have instead
\be
F_0(\eta) = \int^{\sqrt{1-\eta}}_{0}
\frac{u^2du}{1-u^2}\int_u^1\frac{M_0(u)du}{u^2}
\ee
and
\be
F_1(\eta) = \int^{\sqrt{1-\eta}}_{0}\frac{u^2du}{(1-u^2)^2}
\int^u_1\frac{M_1(u)(1-u^2)du}{u^2}
\ee

\subsection{The first terms of the $a(h)$ expansion}

Now one can easily solve (\ref{NGE}) and get
\be
a(\zeta,\bar\zeta) = \sum_{k=0}^\infty
\Big(h_{k+1}\zeta^{k} + \bar h_{k+1}\bar\zeta^{k}\Big)
\left(\frac{\Big(1+k\sqrt{1-\zeta\bar\zeta}\Big)
\Big(1-\sqrt{1-\zeta\bar\zeta}\Big)^{k}}
{(\zeta\bar\zeta)^{k}}
-\zeta\bar\zeta\right)
+  \\
+ \frac{1}{4}\sum_{k,l=0}^\infty \Big(h_{k+1}h_{l+1}\zeta^{k+l} +
\bar h_{k+1}\bar h_{l+1} \bar\zeta^{k+l}\Big)
\left(\frac{k(k+1)}{(1+\sqrt{1-\zeta\bar\zeta})^{k-1}}
+ \frac{l(l+1)}{(1+\sqrt{1-\zeta\bar\zeta})^{l-1}}
\ -\right. \\ \left.
-\ \frac{k(k+1)+l(l+1)+(k+1)(l+1)(k+l)\sqrt{1-\zeta\bar\zeta}
}
{(1+\sqrt{1-\zeta\bar\zeta})^{k+l}}\right)
+  \\
+ \sum_{k>l}^\infty (h_{k+1}\bar h_{l+1} \zeta^{k-l}
+ \bar h_{k+1} h_{l+1}\bar\zeta^{k-l})b_{kl}(z\bar z) +
\sum_k |h_{k+1}|^2 b_{kk}(z\bar z)
+ O(h^3,h^2\bar h,h\bar h^2,\bar h^3)
\ee
where
\be
b_{kl}(\eta) =
\left(\frac{k(k-1)}{2(1+u)^{k+1}}+\frac{l(l-1)}{2(1+u)^{l+1}}
-1\right)\eta^{l+1}
- \frac{k^2+l^2-2+(k+l)(kl-1)u}{2(1+u)^{k+l}}\,\eta^{l} + \\
+ \frac{(k+l)\big(1+(k-l)u\big)}{2(1+u)^{k-l}}
\ee
for $k\geq l$ and $\eta=1-u^2$.
In particular,
\be
b_{k0}(\eta) = g_{k}(\eta)-\eta = \frac{1+ku}{(1+u)^{k}}-\eta
\ee

\subsection{Expression for $A^{(1|1)}$, $A^{(2|1)}$ and $A^{(3|1)}$}

Now we are ready to present a few first expressions for the minimal area and the corresponding
double loop integral. As it was stated by the initial hypothesis, they coincide:

\be
A^{(1|1)}_{i} = \frac{i(i^2-1)}{6} = D^{(1|1)}_{i}\\
A^{(2|1)}_{ij} = \frac{(i+1)(j+1)}{12}\Big(i^2+j^2 + 3ij -
(i+j)\Big) = D^{(2|1)}_{ij} \label{A31} \ee
\be
A^{(3|1)}_{ijk} =
\frac{(i+1)(j+1)(k+1)}{18}\Big(i^2+j^2+k^2 + 3(ij+jk+ik) -
(i+j+k)\Big) = D^{(3|1)}_{ijk} \label{Vco} \ee

\subsection{Expression for $A^{(2|2)}$}

On the other hand, in this (fourth) order we already have an expression which is {\it different}
for the area and the double integral. Indeed, for $k\leq i,j\leq l$
\be
A_{ij|kl}^{(2|2)} = \delta_{i+j,k+l}
\frac{k+1}{48(i+j-1)(i+j+1)} \times  \\ \times
\Big\{2ij(i^4+5i^3j+8i^2j^2+5ij^3+j^4)+2(i+j)^5
-2(k^2-k+1)i^2j^2 +  \\
+k^2(k^2+k-2)(i^2-ij+j^2)
+(3k^4+3k^3-10k^2+4k-2)ij-k^2(k^2+k-2)-  \\
-\frac{1}{i+j}\Big(
2(k^3+k^2-2k+2)(i^4+j^4)
+(7k^3+9k^2-16k+16)ij(i^2+j^2)
+(9k^3+15k^2-24k+24)i^2j^2 -  \\
-2(k^3+k^2-2k+1)(i^2+j^2)
-(5k^3+3k^2-8k+4)ij\Big) \Big\}
\ \ \times h_{i+1}h_{j+1}\bar h_{k+1}\bar h_{l+1}
\ee
while
\be
D^{(2|2)}_{ij|kl} = \delta_{i+j,k+l}
\frac{1}{24}\Big((i+1)(j+1)(k+1)(i^2+3ij+j^2-i-j) -  \\
- (i+j+2)(k+2)(k+1)k(k-1)
+ \frac{3}{5}(k+3)(k+2)(k+1)k(k-1)\Big)
\ \ \times h_{i+1}h_{j+1}\bar h_{k+1}\bar h_{l+1}
\ee
These expressions do not coincide unless one of the indices is equal to 1, i.e. the expansion
contains $h_2$ or $\bar h_2$. In this latter case $A$ and $D$ coincide, as a corollary of the
conformal invariance we discuss in the next section.

Note that these formulas at particular values of indices read as
\be
A_{ij|ij}^{(2|2)} =
\frac{i+1}{48(i+j-1)(i+j)(i+j+1)}\Big(
i(-i^6 -2i^5j +4i^4j^2 +18i^3j^3 +24i^2j^4 +12ij^5 +2j^6) +\\
+(i^6 +6i^5j +20i^4j^2 +34i^3j^3 +28i^2j^4 +12ij^5 +2j^6)
+i(3i^4 +10i^3j +14i^2j^2 +12ij^3 +4j^4)-  \\
-(3i^4 +10i^3j +18i^2j^2 +16ij^3 +4j^4)
-2i(i^2  +4ij + +3j^2)
+2(i+j)^2
\Big)\\
\ \ \times |h_{i+1}|^2|h_{j+1}|^2,\ \ \ \ i\leq j
\ee
and
\be
A^{(2|2)}_{ii|ii} =
\frac{i(i+1)(57i^5+103i^4+43i^3-51i^2-16i+8)}
{96(2i-1)(2i+1)} \ \  \times |h_{i+1}|^4
\ee
while
\be
D^{(2|2)}_{ii|ii} =
\frac{i(i+1)(9i^3-16i^2+9i-4)}{60} \ \  \times |h_{i+1}|^4
\ee
In the following section we are going to describe how the conformal invariance governs this discrepancy.

\section{Conformal invariance}
The structure of $AdS_5$ implies the conformal invariance with generators that form the $SL(2)$-algebra.
In terms of the $h$-parameters calculated in \cite{IMM2} this generators represent a set of
first Virasoro-like constraints
\be
\hat{J}_{-}=\frac{\pr}{\pr h_0}\\
\hat{J}_0=\frac{\pr}{\pr h_1}+\sum\lm_{k=0}^{\infty}h_k\frac{\pr }{\pr h_k}\\
\hat{J}_{+}=\frac{\pr}{\pr h_2}+2\sum_{k=0}^{\infty} h_k \frac{\pr}{\pr h_{k+1}}+\sum_{k,m=0}^{\infty}h_k h_m \frac{\pr}{\pr h_{k+m}}
\ee
Since the minimal area and the double contour integral are equivalent up to the third order,
it deserves investigating the both sides from this point of view.

\subsection{Invariance of the double contour integral}

Consider the action of the global transformation on the double loop integral.
Under the fractional linear $SL(2,R)$ transformation
$z \rightarrow \frac{az+b}{cz+d}$, the ingredients of the
{\it planar} double-loop integral
change as follows
$$
dz_1 d\bar z_2 \rightarrow \frac{dz_1d\bar z_2}
{(cz_1+d)^2(c\bar z_2+d)^2}, \ \ \ \ \
d\bar z_1 dz_2 \rightarrow \frac{d\bar z_1dz_2}
{(c\bar z_1+d)^2(cz_2+d)^2}, \ \ \ \ \
|z_1-z_2|^{-2} \rightarrow |cz_1+d|^2|cz_2+d|^2|z_1-z_2|^{-2},
$$
so that
\be
D \ \stackrel{(\ref{DLI})}{=} \ \frac{1}{2}
\oint\oint \frac{dz_1 d\bar z_2 + d\bar z_1 dz_2}{|z_1-z_2|^2}
\ \longrightarrow \   \\ \longrightarrow \
\frac{1}{2}
\oint\oint \frac{dz_1 d\bar z_2}{|z_1-z_2|^2}\cdot
\frac{c\bar z_1+d}{cz_1+d}\cdot\frac{cz_2+d}{c\bar z_2+d} +
\frac{1}{2}
\oint\oint \frac{d\bar z_1 dz_2}{|z_1-z_2|^2}\cdot
\frac{cz_1+d}{c\bar z_1+d}\cdot\frac{c\bar z_2+d}{cz_2+d}
\ee
It is not very obvious from this formula that $D$
is invariant. However, it actually is.
The simplest way to see this is to look at an infinitesimal
$SL(2,R)$ transformation, picking up the contribution
to the variation of $D$, which is linear in  $d-1$ and $c$.
Then, the variation of $D$ is
\be
\delta D = -\frac{c}{2}\oint\oint
\left(\frac{1}{z_1-z_2} - \frac{1}{\overline{z_1-z_2}}\right)
\Big(d\bar z_1 dz_2 - dz_1d\bar z_2\Big) \
 = c \oint\oint  \frac{y\left(dydX-dxdY\right)}{x^2 +y^2},
\ee
where we denoted $z_1, z_2$ as $z_{1,2} = x_{1,2}+iy_{1,2}$
and $x=x_1-x_2$, $y=y_1-y_2$, $X=x_1+x_2$, $Y=y_1+y_2$.
The integrand is then a total derivative.
In fact,  the above integrals are divergent and
formal manipulations make sense only after regularization,
which, however, breaks the invariance and {\it an anomaly}
 shows up when the regularization is removed:
\be
\delta D = \delta \left(\frac{1}{2}
\oint\oint \frac{dz_1 d\bar z_2 + d\bar z_1 dz_2}
{|z_1-z_2|^2+\lambda^2} \right) =
\frac{\pi}{4\lambda} \delta L
\ee
where $L=L_\Pi$ is the length of the curve $\Pi$.
However, $D^{reg} = D - \pi\frac{L}{4\lambda}$ remains fully
invariant \cite{IMM2}.

\subsection{Invariance of the minimal area}
The initial expression for the minimal area (\ref{SNG}) is obviously invariant under the action
of such generators because of its construction. Introducing the $\mu$-regularization seems to
break this symmetry. Actually, if one considers the action of generators which transforms a
planar contour into another planar contour
\be
r\rightarrow r(1-\gamma \bar z -\bar \gamma z)\\
z\rightarrow z(1-\bar \gamma z)+\gamma r^2\\
\bar z\rightarrow \bar z(1-\gamma \bar z)+\bar \gamma r^2
\ee
the $\mu$-regularization obviously breaks the invariance (in terms of (\ref{plan-NGA}))
\be
\int d^2z \mathcal{L}(z,\bar z,r)\rightarrow \int d^2 z \mathcal{L}(z,\bar z,r)
\left(1-\mu^2\frac{\gamma \bar z+\bar\gamma z}{r^2+\mu^2}\right)
\ee
After regularization, the status of the area invariance is similar to the status of the double
loop integral: on one hand, the expression for the area behaves as $\sim L/\mu+\ldots$,
the violation is proportional to $\mu^2$ and it does not contribute to the regularized part
of the area; on the other hand, the additional divergency arises in the denominator, and this can
cause some troubles.

\bigskip

An explicit calculation of the minimal area in h-terms shows that the conformal invariance
persists. Moreover, it seems to be deeply related to the equality of terms that contain
$h_2$ or $\bar h_2$ at the fourth order of the both sides of (\ref{AMh}).
The invariance arguments can be directly applied to the fifth order contributions into the area,
the corresponding term can be written as follows
\be
A^{(5)}=\frac{3}{2}\sum\lm_{i,j,k,l,m}\left(X_{ijk;lm} h_{i+1} h_{j+1} h_{k+1} \bar h_{l+1} \bar h_{m+1}+Y_{ijkl;m} h_{i+1} h_{j+1} h_{k+1} h_{l+1}\bar h_{m+1}\right)+h.c.
\ee
with the additional condition
\be
X_{ijk;lm}\sim \delta_{i+j+k,l+m}\qquad Y_{ijkl;m}\sim \delta_{i+j+k+l,m}
\ee
One can easily construct a relation which must be satisfied
\be
3!A^{(5)}_{1jk;lm}-2(2! A^{(4)}_{j(k+1);lm}+2!A^{(4)}_{(j+1)k;lm})+2A^{(3)}_{j+k+1;lm}=0
\ee
This relation between coefficients is not universal, it depends on number of coinciding indices in
various terms. To present a dramatic example, consider one of such relations
\be
3A^{(5)}_{112;22}-2(A^{(4)}_{13;22}+A^{(4)}_{22;22})+A^{(3)}_{4;22}=0
\ee
(note that $A^{(5)}_{22;22}$ does not coincide with $D^{(5)}_{22;22}$.)
This relation allows one to anticipate, e.g., the coefficient
$A^{(5)}_{112;22}=\frac{419}{30}$. One can check that this coefficient is, indeed, given by this
number.

\bigskip

In fact, the conformal invariance allows one to fix numerous terms at the both sides.
Consider, for instance, formula (\ref{Vco}). One can expect it is a symmetric
polynomial of degree three in all indices. Actually one knows
it is zero if any one of the indices is
equal to minus one (which corresponds to presence of $h_0$), just since the generator $\hat J_-$
should cancel it due to the conformal invariance.
Therefore, a generic form of this expression is
\be
A^{(3|1)}=\alpha(i+1)(j+1)(k+1)(i^2+j^2+k^2+\beta(ij+jk+ik)+\gamma(i+j+k)+\delta)
\ee
Further, the conformal invariance gives two more equations for $k=0$ and $k=1$:
\be
\alpha(i^2+j^2+\beta ij+\gamma(i+j)+\delta)=\frac{1}{18}(i^2+j^2+3ij-(i+j))\\
2\alpha(i^2+j^2+1+\beta(ij+j+i)+\gamma(i+j+1)+\delta)=\frac{2}{18}(i^2+j^2+3ij+2(i+j))
\ee
These equations allow one to unambiguously define all the coefficients and to restore
the expression not only for $A^{(1...3|1)}$, but for all $A^{(n|1)}$ as well.
Hence, the assumptions of the
polynomial structure and of the conformal invariance of the
regularized minimal area and the double loop integral allow one to claim that the
$h$-linear terms (or $\bar h$-linear ones) in the both sides of (\ref{AMh}) coincide with
each other. Note, however, that the assumption of polynomial structure fails for $A^{(n|m)}$
with min$(m,n)\ge 2$ (though remains true for $D^{(n|m)}$).

\subsection{Towards an explicit "non-perturbative" derivation of the area}

Since the Alday-Maldacena conjecture fails, one has to correct it in some way.
At the moment, one can foresee, at least, what could be elementary building blocks for corrections.

Indeed, the consideration above establishes that $A^{(\cdot|1)}=D^{(\cdot|1)}$.
Then, as in section 4.2.3, one can sum up the $\bar h$-linear terms of the {\it minimal area}
\be
A^{(\cdot|1)} \equiv
\sum_{m=1}^\infty A^{(m|1)}_{k_1\ldots k_m|k_1+\ldots+k_m}
h_{k_1+1}\ldots h_{k_m+1}\bar h_{k_1+\ldots+k_m+1} =\frac{1}{6}
\oint \bar h(\bar \zeta)S_\zeta(z)\zeta^2 d\zeta
\ee
This does not come as a surprise that the result is expressed through the conformal
invariant Schwarzian derivative. Moreover, if one anticipates that it enters linearly, the answer
is completely fixed up to a coefficient. This explains why $A^{(\cdot|1)}=D^{(\cdot|1)}$.

To construct
further $A$, one has to introduce higher conformal structures into the game.
One can directly make certain that the following structures are invariant under the rational
transformation, which reflects action of the conformal $SL(2)$-algebra:
\be
S(z) = \frac{z'''}{z'} - \frac{3}{2}\left(\frac{z''}{z'}\right)^2
\ee
\be
S_2(z_1,z_2)\sim\frac{dz_1 dz_2}{(z_1-z_2)^2} =
\frac{d\zeta_1d\zeta_2}{(\zeta_1-\zeta_2)^2}
\times\\ \times
\left(1 + \sum_k kh_k\zeta_1^{k-1}\right)
\left(1 + \sum_k kh_k\zeta_2^{k-1}\right)
\left\{1 + \sum_{m=1}^\infty (-)^m(m+1)
\left(h_k\frac{\zeta_1^k-\zeta_2^k}{\zeta_1-\zeta_2}\right)^m\right\}
\ee
\be
S_3(z_1,z_2,z_3)\sim\frac{dz_1 dz_2 dz_3}{(z_1-z_2)(z_2-z_3)(z_3-z_1)} =
\frac{d\zeta_1d\zeta_2d\zeta_3}
{(\zeta_1-\zeta_2)(\zeta_2-\zeta_3)(\zeta_3-\zeta_1)}\times\\
\times\left(1 + \sum_k kh_k\zeta_1^{k-1}\right)
\left(1 + \sum_k kh_k\zeta_2^{k-1}\right)
\left(1 + \sum_k kh_k\zeta_3^{k-1}\right)
\cdot
\ee
These expressions should serve as building blocks for higher $A$, which becomes
different from $D$ already in the $\bar h^2$-term, i.e. at the
level of non-perturbative $A^{(\cdot|2)}$. However, a close inspection of this
difference as well as of the invariants is beyond the scope of the present paper.

\section{Non-planar case}
\subsection{Equations}

To go beyond the planar case, i.e. to lift the last requirement in (\ref{circ}),
we switch on one of the coordinate fields
that was put zero before (for the sake of definiteness, we choose $y_0$).
Thus, we add its contributions to the area and to the loop integral
\be
A=\int \frac{\sqrt{-8\pr r \bar \pr r \pr y_0 \bar \pr y_0 +4 \pr r \bar \pr r \pr H \bpr \bar H-4 \pr y_0\bpr y_0\pr H\bpr\bar H+4(\pr r)^2(\bpr y_0)^2+4(\bpr r)^2(\pr y_0)^2+(\pr H \bpr \bar H)^2}}{r^2+\mu^2}\\
D=\frac{1}{4}\oint\oint \frac{\frac{1}{2}(dz d\bar z'+dz'd\bar z)-dy_0 dy_0'}{(z-z')(\bar z-\bar z')-(y_0-y_0')^2}
\ee
As previously, we use the gauge invariance to fix the fields $y_1$ and $y_2$
as in (\ref{RiemPar}). The corresponding Euler-Lagrange equations is a set of two non-linear
equations for the fields $r$ and $y_0$. In contrast to the boundary condition
$\left.r\right|_{\zeta\bar\zeta=1}=0$, we consider for $y_0$
\be\label{npbc}
\left.y_0(\zeta,\bar \zeta)\right|_{\zeta\bar\zeta=1}=\sum\lm_{k=0}^{\infty} q_k \zeta^k+\sum\lm_{k=1}^{\infty}q_{-k}\bar\zeta^{k}
\ee
which is nothing but the Fourier expansion of an arbitrary boundary condition.
We again develop an iterative procedure, looking now for formal expansion in powers of both the
$h$- and $q$-parameters. Beyond the boundary (in the bulk) one has
\be
r(\zeta,\bar\zeta)=\sqrt{1-\zeta\bar\zeta+a^{(1)}(\zeta,\bar\zeta)+a^{(2)}(\zeta,\bar\zeta)+\ldots}\\
y_0(\zeta,\bar\zeta)=b^{(1)}(\zeta,\bar\zeta)+b^{(2)}(\zeta,\bar\zeta)+\ldots
\ee
and the NG equations convert into
\be
\Delta_{NG}(a^{(1)}(\zeta,\bar \zeta)+h(\zeta)\bar\zeta+\bar h(\bar\zeta)\zeta)=0\\
\label{b-eq}\Delta_{NG}b^{(1)}(\zeta,\bar \zeta)=0\\
\Delta_{NG}a^{(k)}(\zeta,\bar \zeta)=-R^{(k)}(a,b;h,\bar h)\\
\Delta_{NG}b^{(k)}(\zeta,\bar \zeta)=-Q^{(k)}(a,b;h,\bar h)
\ee
One can easily construct the solution of the second equation in (\ref{b-eq})
which satisfies the boundary conditions
(\ref{npbc})
\be
b^{(1)}(\zeta,\bar\zeta)=\sum\lm_{k=0}^{\infty} q_k \zeta^k g_{k}(\zeta\bar\zeta)+\sum\lm_{k=1}^{\infty}q_{-k}\bar\zeta^{k}g_k(\zeta\bar\zeta)
\ee
Then, all other iterations must satisfy the trivial boundary condition
$\left.b^{(k)}\right|_{\zeta\bar\zeta=1}=0,\quad k\geq 2$.

\subsection{Explicit results}

The first $q$-dependent contributions to the regularized area and to the double loop integral
are
\be
\frac{A}{3\pi}=\frac{D}{\pi^2}=\sum\lm_{k=1}^{\infty}\frac{2}{3}k(k^2-1)q_kq_{-k}+\sum\lm_{i,j\geq 1}\frac{ij(i^2+j^2+3(i+j)-2)}{6}\left(q_i q_j\bar h_{i+j+1}+q_{-i}q_{-j}h_{i+j+1}\right)-\\
-\frac{(i+1)j(i^2+2j^2+3ij+2i-2)}{3}\left(h_{i+1} q_j q_{-i-j}+\bar h_{i+1} q_{-j}q_{i+j}\right)
\ee
Therefore, up to the third order, $A$ and $D$ coincide.

In the fourth order, the situation is much similar to the planar case.
For instance, the both expressions are of the form
\be
A,D\sim T_{ij;kl}q_i q_j q_{-k} q_{-l} \delta_{i+j,k+l}+Y_{ijk} (q_i q_j q_k q_{-i-j-k}+q_{-i}q_{-j}q_{-k}q_{i+j+k})
\ee
where the coefficients $T_{ij;kl}$ are generally different, while
\be
Y_{ijk}\sim ijk\left(\frac{2}{3}(i^2+j^2+k^2)+(ij+kj+jk)-2\right)
\ee
coincide for the both sides. This expression must be fixed by the conformal invariance as
in the planar case.

\subsection{Conformal invariance in the non-planar case}

$AdS_{d+1}$ space can be considered as a quadric (hyperboloid)
in $R^{d+2}$, defined by
\be
Y_+Y_- + \vec Y^2 = 1,
\label{embed}
\ee
which can be conveniently parameterized by
\be
\vec y = \frac{\vec Y}{Y_+}, \ \ \ r=\frac{1}{Y_+},  \\
\vec Y = \frac{\vec y}{r}, \ \ \ Y_+ = \frac{1}{r}, \ \ \
Y_- = \frac{r^2-\vec y^2}{r}
\ee
The metric on $AdS_{d+1}$, induced by this embedding from the
flat one, has the Poincar\'e form
\be
dY_+dY_- + d\vec Y^2 = \frac{d\vec y^2 - dr^2}{r^2}
\label{AdSmet}
\ee
The symmetry transformations from $SO(d,2)$ act linearly
on the $(d+2)$-component vector $(Y_+,Y_-,\vec Y)$,
\be
\delta\left(\begin{array}{c} Y_- \\ Y_+ \\ \vec Y \end{array}\right)=
\left(\begin{array}{ccc} b & 0 & 2\vec\beta\\ 0 & -b& 2\vec\gamma
\\ -\vec\gamma & -\vec\beta & B \end{array}\right)
\left(\begin{array}{c} Y_- \\ Y_+ \\ \vec Y \end{array}\right)
\ee
Consider the action of the $\vec{\gamma}$-generator on $AdS_5$:
\be
\delta Y_-=0, \quad \delta Y_+=2(\vec{\gamma}\,\vec{Y}),\quad \delta \vec{Y}=-\vec{\gamma}\,Y_-
\ee
Expressing $Y$'s via the coordinates on $AdS_5$, one can immediately derive the transformations
of these latter:
\be
\delta r=-2(\vec{\gamma}\,\vec{y})r\\
\delta \vec{y}=-2\vec{y}\,(\vec{\gamma}\,\vec{y})+\vec{\gamma}\,\vec{y}\,^2-\vec{\gamma}r^2
\ee
This means that the two planar generators can be realized at the boundary ($r=0$) as
\be
\hat{\gamma}_1=2y_0y_1\frac{\pr}{\pr y_0}+(y_1^2+y_0^2-y_2^2)\frac{\pr}{\pr y_1}+2y_1 y_2\frac{\pr}{\pr y_2}\\
\hat{\gamma}_2=2y_0y_2\frac{\pr}{\pr y_0}+2y_1 y_2\frac{\pr}{\pr y_1}+(y_2^2+y_0^2-y_1^2)\frac{\pr}{\pr y_2}
\ee
or, equivalently,
\be
\hat{\bar{\gamma}}=\hat{\gamma}_1+i\hat{\gamma}_2=2y_0(y_1+i y_2)\frac{\pr}{\pr y_0}+y_0^2\left(\frac{\pr}{\pr y_1}+i\frac{\pr}{\pr y_2}\right)+\left(y_1^2-y_2^2+2iy_1y_2\right)\frac{\pr}{\pr y_1}+
\\+\left(2y_1y_2+i y_2^2-i y_1^2\right)\frac{\pr}{\pr y_2}=2y_0z \frac{\pr}{\pr y_0}+2y_0^2\frac{\pr}{\pr \bar z}+z^2\left(\frac{\pr}{\pr y_1}-i\frac{\pr}{\pr y_2}\right)=\\
=2\left(y_0 z\frac{\pr}{\pr y_0}+y_0^2\frac{\pr}{\pr \bar z}+z^2\frac{\pr }{\pr z}\right)
\ee

Let us now check the invariance of the double contour integral \be
D=\oint\oint\frac{(dy dy')}{(y-y')^2+\lambda^2} \ee w.r.t. the
action of the generator $\vec{\hat{\gamma}}$: \be \delta
D=2\oint\oint\frac{(dy d\delta
y')}{(y-y')^2+\lambda^2}-2\oint\oint\frac{(dy d
y')(y-y',\delta(y-y'))}{\left[(y-y')^2+\lambda^2\right]^2}=
\\=
2\oint\oint\frac{dy(-2dy'(\gamma y')-2y'(\gamma dy')+2\gamma(y'\,dy'))}{(y-y')^2+\lambda^2}+2\oint\oint\frac{(y-y')^2(y+y',\gamma)}{\left[(y-y')^2+\lambda^2\right]^2}=\\
=-4\lambda^2\oint\oint\frac{(dy\,dy')(y'\gamma)}{\left[(y-y')^2+\lambda^2\right]^2}+4\oint\oint\frac{(\gamma dy')((y-y')dy)}{(y-y')^2+\lambda^2}=\\
=-4\lambda^2\oint\oint\frac{(dy\,dy')(y'\gamma)}{\left[(y-y')^2+\lambda^2\right]^2}+2\oint
(\gamma dy')\oint d\log \left[(y-y')^2+\lambda^2\right] \ee The
second term is a total derivative and can be neglected. The first
term seems to tend to zero since $\lambda\rightarrow 0$, on the
other hand there is an additional divergency in the denominator and
the invariance can not be claimed as obvious.

Similarly, other generators of the conformal transformations are (the $SO(d)$ rotations
omitted):
\be\label{generators}
\hat\beta_0: \hfill\frac{\partial}{\partial y_0}\ , \\
\hat\beta:  \hfill\frac{\partial}{\partial z}\ , \\
\hat{\bar\beta}: \hfill \frac{\partial}{\partial\bar z}\ ,  \\
\hat b: \hfill y_0\frac{\partial}{\partial y_0} + z\frac{\partial}{\partial z}
+ \bar z\frac{\partial}{\partial\bar z}\ , \\
\hat\gamma_0: \hfill \phantom{acd}(z\bar z+y_0^2)\frac{\partial}{\partial y_0}
+2y_0\left(z\frac{\partial}{\partial z} +
\bar z\frac{\partial}{\partial\bar z}\right), \\
\hat\gamma: \hfill y_0\bar z\frac{\partial}{\partial y_0} +
y_0^2\frac{\partial}{\partial z}
+ \bar z^2\frac{\partial}{\partial\bar z}\ , \\
\hat{\bar\gamma}: \hfill y_0z\frac{\partial}{\partial y_0} +
z^2\frac{\partial}{\partial z} +
\underline{y_0^2\frac{\partial}{\partial \bar z}} \ee

One can represent these generators in the $h$,$q$-variables using
the following relations
\be
\frac{\pr \mathcal{F}}{\pr h_k}=\oint ds\,\zeta^k \frac{\delta \mathcal{F}}{\delta z},\quad
\ \ \ \ \ \ \frac{\pr \mathcal{F}}{\pr q_k}=\oint ds\, \zeta^k \frac{\delta \mathcal{F}}{\delta y_0}
\ee
Then, one obtains
\be\label{gen_ind}
\hat b=\frac{\pr}{\pr h_1}+\frac{\pr }{\pr \bar h_1}+\sum\lm_{k=1}^{\infty} \bar h_k \frac{\pr}{\pr \bar h_k}+\sum\lm_{k=-\infty}^{\infty}q_k \frac{\pr }{\pr q_k}\\
\hat {\bar \gamma}=\frac{\pr}{\pr h_2}+2\sum\lm_{k=1}^{\infty}h_k\frac{\pr}{\pr h_{k+1}}+\sum\lm_{k,m\geq 1}h_k h_m\frac{\pr}{\pr h_{k+m}}+\sum\lm_{k=-\infty}^{\infty}q_k \frac{\pr}{\pr q_{k+1}}+\sum\lm_{k,m}q_k h_m \frac{\pr}{\pr q_{k+m}}+\\
+\sum\lm_{k,m\geq 0} q_{-k}q_{-m}\frac{\pr}{\pr \bar h_{k+m}}+2\sum\lm_{k\geq m}q_mq_{-k}\frac{\pr}{\pr \bar h_{k-m}}
\ee

Now let us see how the conformal invariance works in the non-planar
case. To this end, consider the following terms
\be\label{MR} \sum_k
M_k q_kq_{-k} + \sum_{i,j}R_{ij}(h_{i+1}q_jq_{-i-j} + \bar
h_{i+1}q_{-j}q_{i+j})+ \sum A_k h_{k+1}\bar h_{k+1} + \sum_{k,m}
B_{k,m}q_{-k-m-1}q_{m}h_{k+1}h_2
\ee
They should (and do) satisfy
the non-planar-non-planar relations
\be\label{104}
2R_{0j} + 2M_{j}=0,  \\
R_{1j} + M_{j+1} + M_j=0
\ee
where 2 at the l.h.s. of the first formula comes from
the action of $\partial/\partial h_1$ and $\partial/\partial\bar h_1$,
which are both present in ${\hat b}$.

Less trivial example arises when one acts with the generators onto (\ref{MR}) and compare coefficients
in front of $q_{-k-m}q_{m}h_k$. This leads to the following relation for the coefficients
\be\label{confexp}
B_{k-1,m}+2 R_{k,m}+R_{k-1,m}+R_{k-1,m+1}+M_{k+m}+M_m+2A_{k-1}=0
\ee
The expression for $B$ can be explicitly found and reads
\be\label{106}
B_{k,m}=\frac{2}{3}m(k+1)(2k^2+6km+7k+4m^2+6m+2)
\ee
Formula (\ref{confexp}) is also checked to be valid.

Note that this derivation of formulas for the conformal generators
is, in fact, a bit naive. In principle, one should take into account
that that the conformal transformations change the gauge-fixing
condition, as one can already see from (\ref{generators}). Indeed,
the generators $\hat \gamma$'s explicitly mix purely holomorphic
$z$-coordinates with $y_0$ which contains anti-holomorphic parts.
Still, eqs.(\ref{104})-(\ref{106}) demonstrate that the generators (\ref{gen_ind})
are adequate, at least, in these examples.

\subsection{Non-planar generalization of Schwarzian}

Now we are going to make an attempt to construct the same relation in invariant terms and
introduce a kind of "non-planar" Schwarzian. To this end, we add to $(z,\bar z)$ the holomorphic
and anti-holomorphic parts of $y_0$, $(y,\bar y)$ so that under the conformal transformation
\be\label{pro}
z\rightarrow \frac{az+b}{cz+d},\ \ \ \ \ \ \  y \rightarrow \frac{y}{cz+d}
\ee
Then
\be
z' \rightarrow \frac{z'}{cz+d},  \\
z'' \rightarrow \frac{z''}{(cz+d)^2} - \frac{2c(z')^2}{(cz+d)^3}, \\
z''' \rightarrow \frac{z'''}{(cz+d)^2}
-\frac{6cz'z''}{(cz+d)^3} + \frac{6c^2(z')^3}{(cz+d)^4}
\ee
and
\be
y' \rightarrow \frac{y'}{cz+d} - \frac{cyz'}{(cz+d)^2}, \\
y'' \rightarrow \frac{y''}{cz+d} - \frac{2cy'z'}{(cz+d)^2}
- \frac{cyz''}{(cz+d)^2} + \frac{2c^2y(z')^2}{(cz+d)^3},  \\
y''' \rightarrow \frac{y'''}{cz+d} - \frac{3cy''z'}{(cz+d)^2}
-\frac{3cy'z''}{(cz+d)^2} + \frac{6c^2y'(z')^2}{(cz+d)^3}
-\frac{cyz'''}{(cz+d)^2} + \frac{6c^2yz'z''}{(cz+d)^3}
-\frac{6c^3y(z')^3}{(cz+d)^4}
\ee
It follows that
\be
S(z) = \frac{z'''}{z'} - \frac{3}{2}\left(\frac{z''}{z'}\right)^2
\rightarrow S(z)
\ee
is invariant, while
\be
\sigma_2 (y,z) = \left(\frac{y''}{z'} - \frac{y'z''}{(z')^2}\right)
\rightarrow
(cz+d)\sigma_2 (y,z)
\ee
is projective invariant, as well as
\be\label{boxed}
\sigma_3 (y,z) = \left(\frac{y'''}{z'} - \frac{3}{2}\frac{y''z''}{(z')^2}
- \boxed{1}\frac{y'z'''}{(z')^2} + \boxed{\frac{3}{2}}\frac{y'(z'')^2}{(z')^3}\right)
\rightarrow (cz+d) \sigma_3(y,z)
\ee
In a little more detail,
\be
\frac{y'''}{z'} - \frac{3}{2}\frac{y''z''}{(z')^2}
\rightarrow (cz+d)
\left(\frac{y'''}{z'} - \frac{3}{2}\frac{y''z''}{(z')^2}\right)
- cy S(z)
\ee
and this is combined into invariant with $-\frac{y'}{z'}S(z)$
with Schwarzian $S(z)$ invariant and $\frac{y'}{z'} \rightarrow
(cz+d)\frac{y'}{z'} - cy$.

This could imply that the $\bar y$-linear part of the exact answer
is just
\be\label{ndi}
\oint \bar y \Big( \sigma_3(y,z)\zeta^2 + 3 \sigma_2(y,z)\zeta\Big)d\zeta
\ee
However, this is an invariant of transformations (\ref{pro}), i.e. of
\be\label{cgen}
\frac{\partial}{\partial z},  \\
z\frac{\partial}{\partial z} + \frac{1}{2}y\frac{\partial}{\partial y},\\
z^2\frac{\partial}{\partial z} + yz\frac{\partial}{\partial y}
\ee
which are different from (\ref{generators}). Especially important is the lack of the underlined
term from (\ref{generators}) in (\ref{cgen}). Therefore, (\ref{ndi}) is not equal to the
double integral. However, the difference is minor (it is rather simple to verify that (\ref{116})
simulates equation (\ref{confexp})):
\be\label{116}
D=\frac{2}{3}\pi^2\frac{1}{2\pi i}\oint dx \bar y(\bar x)\left(x^2\left(\frac{y'''}{z'}-\frac{3}{2}\frac{y''z''}{z'^2}+\frac{3}{4}\frac{y' z''^2}{z'^3}-\frac{1}{2}\frac{y' z'''}{z'^2}\right)+3x\left(\frac{y''}{z'}-\frac{y'z''}{z'^2}\right)\right)-
\\-\frac{\pi^2}{6}\frac{1}{2\pi i}\oint x^2 dx \bar z(\bar x) S(z)+\ldots
\ee
i.e. the two boxed coefficients in (\ref{boxed}) are actually modified by a factor of $1/2$.
This of course violates the invariance under (\ref{cgen}), but, in fact, this is exactly
what is necessary to ensure the invariance under (\ref{generators}): the change of
coefficients is fully compensated by the change of $\bar z$ by $y^2$ in the term
$\oint\bar z S(z)\zeta^2 d\zeta$ (the missed underlined term in (\ref{cgen})).

\section{Conclusion}

In this paper we applied the constructive approach of
\cite{IMM2},\cite{IMM1}-\cite{M}
and evaluated the regularized area of AdS minimal surface
bounded by a wavy circle at the AdS infinity up to the 4-th
order in deviations $h,\bar h$ from the ideal circle.
We confirm the hypothesis of \cite{IMM2,M} that infinitely many
terms of the form $h^m\bar h$ and $h\bar h^n$ coincide
at the l.h.s. and the r.h.s. of (\ref{AMh}).
However, the $4$-th order involves also the terms $h^2\bar h^2$,
the first ones where the real violation of Alday-Maldacena
hypothesis (\ref{AMh}) can be expected, since these
terms are not controlled by the conformal invariance
\cite{dks,dhks1,dhks3,Komar,IMM2}, supplemented by assumptions about
polynomial dependence of expansion coefficients on their indices.
The deviation from (\ref{AMh}) is indeed observed, moreover, the
assumption of polynomial dependence also turns false for these
terms.

This seems to be in parallel with existing observations
about violation of the BDS conjecture (\ref{BDS}) at $n=6$
\cite{Vo,Koanti}, where the terms, which are not controlled by
the conformal invariance break (\ref{AMh}) and look far more
sophisticated.
An advantage of the wavy lines case is a constructive approach
to the evaluation of minimal areas (still not quite developed
for the finite-$n$ case, for this one should further proceed
along the lines of \cite{IMM1,IM8}).
The disadvantage is the lack of direct contact with
Feynman diagram calculations and with the non-Abelian
Wilson line conjecture at weak coupling.

\bigskip

The wavy line calculation has both beauties and puzzles
of its own. The general structure of the $D^{(m|n)}$
coefficients is fully revealed in s.\ref{Dcalc},
but equally explicit formulas for generic $A^{(m|n)}$
remain to be found, despite a clear recursive procedure
to invert the peculiar Nambu-Goto Laplacian $\Delta_{NG}$
is formulated and applied in s.\ref{Acalc}.
This procedure can be easily computerized,
but, unless properly optimized the calculations
can require a lot of computer time.

On the other hand,
even formulas (\ref{Dgen}) do not possess any
simple representation in terms of differential geometry:
they are not representable as local integrals of
the curvatures of $\Pi$ and its derivatives \cite{IMM2},
thus, their differential geometry meaning remains an open
question.

The same calculation can be (but is not yet) performed for
a wavy deformation of the other solvable example of planar curves:
the two concentric circles \cite{concc}.
Instead we extended the analysis of \cite{IMM2} to
non-planar deformations of the circle.

\bigskip

To conclude, the failure of the Alday-Maldacena conjecture
in its most naive form (\ref{AMh}) is now confirmed both
at finite and infinite $n$.
However, this does not look like an end of the story.
While working out the ways to analyze (\ref{AMh}) we
found that the AdS Plateau problem is not as hopeless as
it looked from the very beginning: one can try and get
concrete and explicit formulas about minimal surfaces and
their regularized areas, which do not at all look
structureless.
This opens a way to systematically investigate the deviations
from (\ref{AMh}), to find hidden symmetries of the both sides
and, finally, to find a modification of the Abelian formula
at the r.h.s. which should serve as a strong-coupling
counterpart of the non-Abelian Wilson average in the weak-coupling
regime and, thus, provide an explicit formulation
of the string/gauge duality for scattering amplitudes.

\section*{Acknowledgements}

Our work is partly supported by the Grant-in-Aid for
Scientific Research 2054278 from the Ministry of Education,
Science and Culture, Japan (H.I.), by NWO project 047.011.2004.026,
by the joint grants 09-01-92440-CE (A.M.'s), 09-02-91005-ANF,
by the Russian President's Grant of
Support for the Scientific Schools NSh-3035.2008.2 and
by Russian Federal Nuclear Energy Agency
(D.G. and A.M.'s), by RFBR grants
07-02-00878 (D.G. and A.Mir.) and 07-02-00645 (A.Mor.).

\newpage

\section*{Appendix: $\kappa$-coefficients in different examples}

We briefly calculate here the coefficient $\kappa$ of proportionality between the
area and the double integral, (\ref{AMh}) and demonstrate that it is typically very close to 1.

For smooth curves the minimal area and the double loop integral are the expressions
of the similar form
\be
A=\kappa_A L+A_{reg}+const_A\\
D=\kappa_D L+D_{reg}+const_B
\ee
Here $L$ is the corresponding length of the boundary where the divergency accumulates.
As we discussed in the paper, $A$ and $D$ can be presented for a smooth curve in the form
\be
A_{\Pi}=\frac{\pi L_{\Pi}}{2\mu} -2\pi-3\pi \frac{1}{6}\frac{1}{2\pi i}\oint dx x^2 \bar h(x) h'''(x)+\ldots \\
D_{\Pi}=\frac{\pi L_{\Pi}}{4\lambda}-\frac{\pi^2}{2}-\pi^2 \frac{1}{6}\frac{1}{2\pi i}\oint dx x^2 \bar h(x) h'''(x) +\ldots
\ee
Therefore, $\kappa_A/\kappa_D=2\lambda/\mu$, while $\kappa_{smooth}$ defined to be
the ratio $A_{reg}/D_{reg}$, is $3/\pi=0.955$ (by modulo difference of $A_{reg}$ by $D_{reg}$
in higher orders and up to inessential numerical additive constants). In fact, to
find $\kappa_A/\kappa_D$, one suffices to consider the case of unit circle. Then,
\be
A=\int\lm_0^1 r dr \int\lm_{0}^{2\pi} d\varphi \frac{1}{\sqrt{1-r^2}(1-r^2+\mu^2)}=2\pi\frac{\arctan\left(\frac{1}{\mu}\right)}{\mu}=\frac{\pi}{2\mu}(2\pi)+\ldots\\
D=\frac{1}{8}\int\int \frac{d\zeta d\bar\zeta'+d\zeta'd\bar\zeta}{(\zeta-\zeta')(\bar \zeta-\bar\zeta')+\lambda^2}=\frac{\pi}{4\lambda}(2\pi)+\ldots
\ee

Another instance is the boundary presented in \cite{AM3}, an infinite strip (two parallel infinite
lines).
We just briefly remind this example following \cite{IMM2}.
The form of the surface can be explicitly found and reads
\be
y(r)=\pm \int\lm_{0}^{r_{max}}\frac{\xi^2 d\xi}{\sqrt{r_{max}^4-\xi^4}}
\ee
A width of the strip (a distance between lines), $a$ is related to $r_{max}$ via
\be
a=\frac{\pi r_{max}}{\sqrt{2}K\left(\frac{1}{\sqrt{2}}\right)}
\ee
where $K$ is the complete elliptic integral of
the second kind. With the $\mu$-regularization the minimal area reads
\be
A=2L r_{max}^2\int\lm_{0}^{r_{max}}\frac{dr}{(r^2+\mu^2)\sqrt{r_{max}^4-r^4}}=
\frac{\sqrt{2}L}{r_{max}}\frac{1}{1+\mu^2}\Pi\left(-\frac{1}{1+\mu^2},\frac{1}{\sqrt{2}}\right)
\ee
where $L$ is the length of the strip, $L\rightarrow\infty$. Thus,
\be
A=\frac{\pi^2 L}{\sqrt{2}K\left(\frac{1}{\sqrt{2}}\right)a}\left(\frac{1}{\mu}-\frac{1}{\sqrt{2}K\left(\frac{1}{\sqrt{2}}\right)}\right)
\ee
The corresponding double loop integral in this case reads
\be
D=\frac{L\pi}{2\lambda}-\pi \frac{L}{2a}
\ee
Therefore, the ration of $A_{reg}$ and $D_{reg}$ is similar
$\kappa_{strip}=4\frac{(2\pi)^2}{\left(\Gamma\left(1/4\right)\right)^4}\approx 0.914$.
Note that in this case $L_{\Pi}=2L+2a\approx 2L$.

At last, consider an example of the rectangular with light-like edges.
The expressions for the minimal area and for the double loop integral are of a slightly different
form in this case so that the divergency is basically due to unsmoothness of the curve.
Then, the dimensional regularization better suits the problem. The minimal area in this
regularization is (up to inessential numerical factors)
\be
A=\frac{2}{\epsilon^2}\left((1+b)^{\epsilon}+(1-b)^{\epsilon}\right)-\frac{1}{4}\log^2 \frac{s}{t}
\ee
where $b$ is the rectangular parameter related to the variables $s,t$ \cite{AM1} via
\be
\frac{(1+b)^2}{(1-b)^2}=\frac{s}{t}
\ee
In this case the ratio of finite parts of the area and the loop integral, $\kappa_{\Box}=1$.
Being a dramatic example, the expression for the area has been calculated in numerical papers
(\cite{AM1,MMT1,Popol} etc). Here we just remind the calculation of the double loop integral
in dimensional regularization (see \cite{MMT1}),
\be
D=\frac{1}{4}\oint\oint\frac{dy^{\mu}dy'_{\mu}}{(y-y')^{(2+\epsilon)}}
\ee
First consider the contribution from angles of the rectangle. We parameterize the two adjacent
edges as
\be
y_0=\frac{a\sqrt{1+b^2}\tanh u}{1+b\tanh u}, \quad y_0=-\frac{a\sqrt{1+b^2}\tanh u'}{1-b\tanh u'}\\
y_1=\frac{a}{1+b\tanh u},\quad y_1=-\frac{a\tanh u'}{1-b\tanh u'}\\
y_2=\frac{a\tanh u}{1+b\tanh u},\quad y_2=\frac{a}{1-b\tanh u'}
\ee
Here $a$ is an arbitrary parameter.
After substitution and change of variables, one can immediately derive
\be
D_{angle}=\frac{1}{4}(b+1)^2\left(\int\lm_{-1}^1\frac{dt}{(1+b t)^{1-\epsilon/2}(1-t)^{1+\epsilon/2}}\right)^2=2^{\epsilon}\frac{(1-b)^{\epsilon}}{\epsilon^2}
\ee
The factor $-2$ is since we change the direction of integration in the second integral
and the contour is passed twice in different integrals. The contribution from the
opposite angles is evidently the same, while the other two angles contribute with the opposite
sign of $b$. The contribution of the opposite sides is convergent and no regularization is needed:
\be
D_{opposite}=-4\frac{1}{8}(1+b^2)\int\lm_{-1}^{1}dt\int\lm_{-1}^{1}dt'\frac{1}{(1+b t)(1-b t')(1-tt')}=
\\=-\frac{\pi^2}{4}-\log^2\left(\frac{1+b}{1-b}\right)=-\frac{\pi^2}{4}-\frac{1}{4}\log^2\frac{s}{t}
\ee
The factor $4$ arises because of 4 opposite edges and corresponding integrations.

\end{document}